\documentclass[pdflatex,sn-mathphys-num]{sn-jnl}

\usepackage{graphicx}%
\usepackage{multirow}%
\usepackage{amsmath,amssymb,amsfonts}%
\usepackage{amsthm}%
\usepackage{mathrsfs}%
\usepackage[title]{appendix}%
\usepackage{xcolor}%
\usepackage{textcomp}%
\usepackage{manyfoot}%
\usepackage{booktabs}%
\usepackage{algorithm}%
\usepackage{algorithmicx}%
\usepackage{algpseudocode}%
\usepackage{listings}%
\usepackage{xspace}
\usepackage{float}
\usepackage{natbib}
\usepackage{courier}
\usepackage{lineno}
\usepackage{makecell}

\newcommand{\nsk}{NeuroSketch\xspace}
\newcommand{\tcn}{ModernTCN\xspace}
\newcommand{\medformer}{Medformer\xspace}
\newcommand{\gru}{MultiResGRU\xspace}

\newcommand{\llmbase}{Qwen2.5-7B-Instruct\xspace}
\newcommand{\llmtranslate}{Qwen2.5-7B-Translate\xspace}
\newcommand{\llmrank}{Qwen2.5-7B-Rank\xspace}
\newcommand{\llmcorrect}{Qwen2.5-7B-Correct\xspace}
\newcommand{\qwenmiddle}{Qwen2.5-72B-Instruct\xspace}
\newcommand{\qwenmax}{Qwen-3 Max\xspace}
\newcommand{\deepseek}{Deepseek-v3.2-exp\xspace}
\newcommand{\doubao}{Doubao-1.6\xspace}
\newcommand{\gpt}{GPT-5-chat-latest\xspace}

\newcommand{\grok}{Grok-4-fast\xspace}
\newcommand{\llama}{Llama-3.1\xspace}

\newcommand{\vpara}[1]{\vspace{0.01in}\noindent\textbf{#1 }}

\theoremstyle{thmstyleone}%
\theoremstyle{thmstyletwo}%

\theoremstyle{thmstylethree}%

\begin{document}

\title[Article Title]{Towards unified brain-to-text decoding across speech production and perception}


\author[1]{\fnm{Zhizhang} \sur{Yuan}}\email{zhizhangyuan@zju.edu.cn}
\equalcont{These authors contributed equally to this work.}

\author*[1]{\fnm{Yang} \sur{Yang}}\email{yangya@zju.edu.cn}
\equalcont{These authors contributed equally to this work.}

\author[1]{\fnm{Gaorui} \sur{Zhang}}\email{gaoruizhang@zju.edu.cn}

\author[2]{\fnm{Baowen} \sur{Cheng}}\email{chengbaowen23@mails.ucas.ac.cn}

\author[3]{\fnm{Zehan} \sur{Wu}}\email{zhwu08@fudan.edu.cn}

\author[3]{\fnm{Yuhao} \sur{Xu}}\email{23211220055@m.fudan.edu.cn}

\author[3]{\fnm{Xiaoying} \sur{Liu}}\email{7631935@qq.com}

\author[3]{\fnm{Liang} \sur{Chen}}\email{hschenliang@fudan.edu.cn}

\author[3]{\fnm{Ying} \sur{Mao}}\email{maoying@fudan.edu.cn}

\author*[2]{\fnm{Meng} \sur{Li}}\email{li.meng@mail.sim.ac.cn}

\affil[1]{\orgdiv{Computer Science and Technology}, \orgname{Zhejiang University}, \orgaddress{\city{Hangzhou}, \state{Zhejiang}, \country{China}}}

\affil[2]{\orgdiv{Shanghai Institute of Microsystem and Information Technology}, \orgname{Chinese Academy of Sciences}, \orgaddress{\city{Shanghai}, \country{China}}}

\affil[2]{\orgdiv{Department of Neurosurgery}, \orgname{Huashan Hospital, Fudan University}, \orgaddress{\city{Shanghai}, \country{China}}}


\abstract{
Speech production and perception constitute two fundamental and distinct modes of human communication. Prior brain-to-text decoding studies have largely focused on a single modality and alphabetic languages.  
Here, we present a unified brain-to-sentence decoding framework for both speech production and perception in Mandarin Chinese. 
The framework exhibits strong generalization ability, enabling sentence-level decoding when trained only on single-character data and supporting characters and syllables unseen during training. 
In addition, it allows direct and controlled comparison of neural dynamics across modalities.
We collected neural data from 12 participants implanted with depth electrodes and achieved full-sentence decoding across multiple participants, with best-case Chinese character error rates of 14.71\% for spoken sentences and 21.80\% for heard sentences.
Mandarin speech is decoded by first classifying syllable components in Hanyu Pinyin, namely initials and finals, from neural signals, followed by a post-trained large language model (LLM) that maps sequences of toneless Pinyin syllables to Chinese sentences.
To enhance LLM decoding, we designed a three-stage post-training and two-stage inference framework based on a 7-billion-parameter LLM, achieving overall performance that exceeds larger commercial LLMs with hundreds of billions of parameters or more.
In addition, several characteristics were observed in Mandarin speech production and perception: speech production involved neural responses across broader cortical regions than auditory perception; channels responsive to both modalities exhibited similar activity patterns, with speech perception showing a temporal delay relative to production; and decoding performance was broadly comparable across hemispheres.
Our work not only establishes the feasibility of a unified decoding framework but also provides insights into the neural characteristics of Mandarin speech production and perception.
These advances contribute to brain-to-text decoding in logosyllabic languages and pave the way toward neural language decoding systems supporting multiple modalities.
}

\maketitle

\section*{Introduction}
\label{introduction}
Language is a uniquely human system for representing and communicating information, expressed through several communicative modalities such as speaking, listening, reading, and writing. Among these modalities, speaking and listening constitute the primary means of everyday communication. 
Previous work has demonstrated that linguistic content can be decoded from brain activity elicited during speaking~\cite{qian2025real,metzger2023high,Willett2023HighPerformance,Chen2024NeuralSpeech,Card2024AccurateRapid,Feng2025AcousticInspired,Zheng2024DuIN,Duraivel2023HighResolution,kunz2025inner,Anumanchipalli2019SpeechSynthesis,Singh2025TransferLearning,Luo2023StableDecoding,moses2021neuroprosthesis} or listening~\cite{Defossez2023DecodingSpeech,Fodor2024TowardsDecoding} by establishing mappings between neural signals and the corresponding linguistic information. Most existing studies focus on one of these modalities, each adopting its own experimental paradigm for data collection and a corresponding decoding pipeline. For data acquisition, many high-accuracy language decoding studies rely on recording methods such as microelectrode arrays~\cite{Card2024AccurateRapid, Willett2023HighPerformance}, which capture local neural activity with extremely high temporal and spatial resolution. Nevertheless, their spatial coverage is limited to highly localized functional areas, which makes multimodal decoding challenging.
Regarding decoding framework design, models have been developed for different tasks, including classifying isolated words~\cite{Zheng2024DuIN}, inferring sentences from acoustic features~\cite{Feng2025AcousticInspired,Card2024AccurateRapid,metzger2023high,Willett2023HighPerformance}, and reconstructing speech waveforms~\cite{metzger2023high,Chen2024NeuralSpeech,Defossez2023DecodingSpeech}. 
Meanwhile, the vast majority of these advances have been achieved in alphabetic language systems~\cite{metzger2023high,Willett2023HighPerformance,Chen2024NeuralSpeech,Card2024AccurateRapid,Duraivel2023HighResolution,kunz2025inner,Anumanchipalli2019SpeechSynthesis,Singh2025TransferLearning,Luo2023StableDecoding,moses2021neuroprosthesis, Defossez2023DecodingSpeech,Fodor2024TowardsDecoding}, such as English and Dutch.
In contrast, brain-to-text decoding studies on logosyllabic languages, most notably Chinese, the language with the largest number of native speakers worldwide, remain comparatively limited in brain-to-text decoding~\cite{qian2025real, Zheng2024DuIN, Feng2025AcousticInspired, Zhang2024BrainToText}.

\begin{figure}[tbp]
   \centering
   \includegraphics[width=1.0\textwidth]{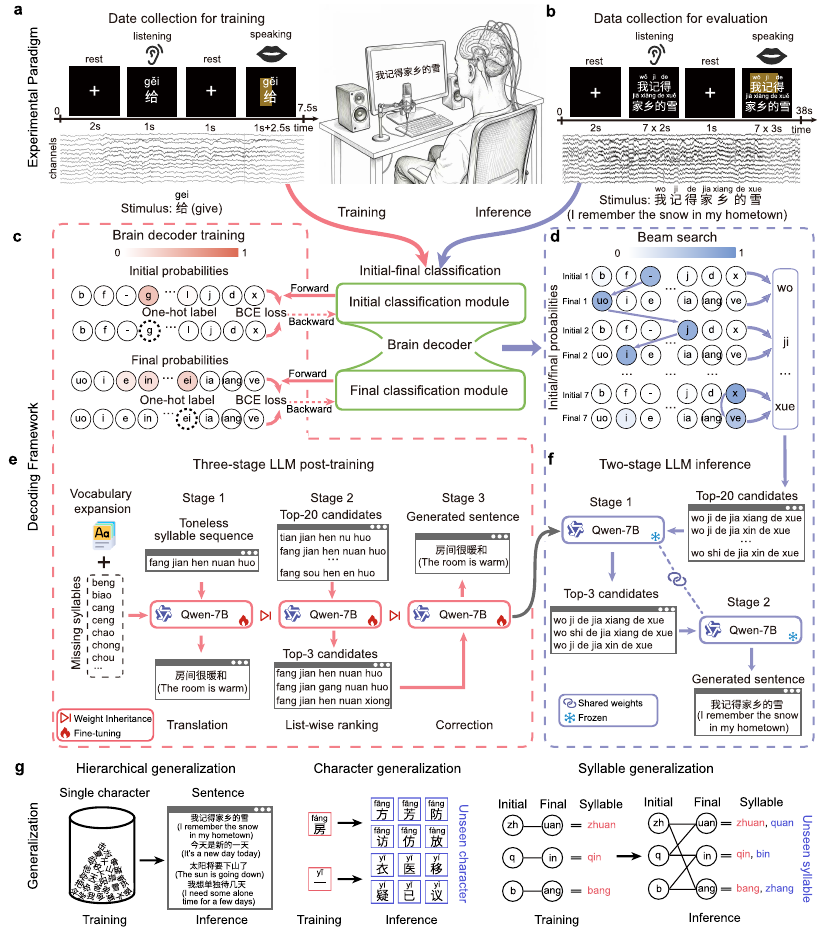}
   \caption{
   \textbf{Overview of our brain-to-sentence framework.}
   \textbf{a,b,} Data acquisition paradigm. 
   \textbf{a,} Task alternating between listening to and speaking individual Chinese characters.
   \textbf{b,} Task alternating between listening to and speaking complete Mandarin sentences.
   \textbf{c,e,} Training pipeline of the decoding framework.
   \textbf{c,} the brain decoder is trained to classify syllable components (initials and finals) from neural signals. We adopted \nsk, a 2D-CNN-based neural decoder developed in our concurrent work. 
   \textbf{e,} The LLM is post-trained to generate full sentences from syllable sequences.
   \textbf{d,f,} Inference pipeline of the decoding framework.
   \textbf{d,} Beam search is applied to the probabilistic outputs of the brain decoder to generate multiple syllable sequence candidates, from which the top-20 candidates are retained.
   \textbf{f,} The top-20 candidates are fed into the post-trained LLM, which performs a two-stage inference process to produce the final decoded sentence.
   \textbf{g,} The proposed framework exhibits hierarchical, character-level, and syllable-level generalization abilities.
   } 
   \label{fig:main_fig}
\end{figure}

Addressing this gap, we present a unified brain-to-sentence decoding framework for Mandarin Chinese that operates seamlessly across both speaking and listening modalities. 
We employed stereoelectroencephalography (sEEG) by implanting multiple depth electrodes across various functional brain regions in each participant, establishing the foundation for multimodal neural decoding.
Our approach unifies the two modalities from the level of behavioral paradigms through the entire decoding pipeline, enabling not only sentence decoding under either modality but also a direct comparison of the neural dynamics evoked by speech production and perception.
Mandarin Chinese employs logographic characters, which number in the tens of thousands~\cite{hanyu_dazidian}, making direct character-level decoding from neural activity impractical.
We instead leverage Hanyu Pinyin, the standardized phonetic system composed of initials, finals, and tones~\cite{Li1998ContextEffects}, covering over 1200 possible tonal syllables, where each syllable maps to multiple Chinese characters.
Our decoding framework utilizes only initials and finals to form toneless syllables, as tone decoding was less robust than initial and final decoding, and incorporating tone information led to suboptimal decoding performance, as examined in subsequent analyses.
The decoding pipeline consists of three stages. At first, the initial and final of each Chinese character are classified from neural signals. Then, the resulting classification probabilities are searched to generate multiple candidate syllable sequences. Finally, the correct Chinese sentence is inferred based on these candidates.
The framework exhibits generalization ability (shown in Fig.~\ref{fig:main_fig}g). First, it provides hierarchical generalization, as training solely on neural responses to isolated spoken or heard characters is sufficient for decoding full sentences. Second, it achieves character generalization, as the decoded sentences may include Chinese characters that never appeared during training. Third, it enables syllable generalization, allowing the model to decode Chinese pinyin syllables that were not present in the training set.
Together, these properties establish a general, modality-unified, and linguistically grounded decoding framework for Mandarin Chinese.

A major technical challenge in the decoding process arises in the final stage, which aims to infer the correct sentence from the searched candidate set. 
In alphabetic languages such as English, acoustic features are represented at the phoneme level, and a phoneme sequence strongly constrains the lexicon, typically mapping to a unique word or only a few candidate words. 
As a result, in English speech decoding, the search process effectively performs a phoneme-to-word mapping.
In contrast, decoding from a toneless syllable to a logographic character in Mandarin Chinese involves a two-step one-to-many mapping: first from the toneless syllable to the tonal syllable, and then from the tonal syllable to the logographic character, resulting in a single toneless syllable potentially corresponding to dozens of different characters. 
Similarly, at the word level, toneless syllable sequences still provide insufficient contextual constraints, resulting in persistent one-to-many mappings analogous to those observed at the character level.
Consequently, character- or word-level mappings are inherently highly ambiguous, introducing cumulative errors that may mislead subsequent correction steps.

Although toneless syllables are highly ambiguous in isolation, sentence context often provides strong constraints.
Therefore, we designed an end-to-end approach that leverages large language models (LLMs) to decode toneless syllable sequences into Chinese sentences directly, eliminating intermediate translation steps, while enabling the model to exploit contextual information within and across sequences.
Nevertheless, LLMs with several billion parameters struggle to accomplish this task effectively, likely due to insufficient training on this modality (toneless syllable sequences) and closely related inference tasks.
Although commercial LLMs with hundreds of billions to trillions of parameters demonstrate better performance, their inference incurs substantial computational costs and is impractical for local deployment.
To address this issue, we developed a three-stage post-training and two-stage inference scheme based on a 7-billion-parameter model (shown in Fig.~\ref{fig:main_fig}e,f), which demonstrated superior performance to some commercial LLMs on this task.
Our results demonstrate that, in neural language decoding tasks, continued post-training enables LLMs to process diverse input types and tackle more complex decoding tasks, moving beyond their traditional role of merely correcting or refining final decoded sentences.

We collected neural data from 12 participants implanted with depth electrodes, across whom we achieved reliable decoding of speech at multiple levels.
Initial and final decoding performance was significantly above chance level for all participants in the speaking task (mean initial accuracy = 59.54\%; mean final accuracy = 50.17\%) and for 10 participants in the listening task (mean initial accuracy = 58.92\%; mean final accuracy = 48.05\%).
Building on these results, sentence-level decoding results showed that 6 participants in the speaking task and 5 participants in the listening task achieved average character error rates (CERs) below 50\% (mean speaking CER = 31.52\%; mean listening CER = 37.28\%), with the best CERs reaching 14.71\% and 21.80\%, respectively. Among these, 4 participants demonstrated reliable sentence decoding across both modalities (mean speaking CER = 32.15\%; mean listening CER = 36.80\%).
Furthermore, we identified several additional observations characterizing the neural features of speech production and perception. 
First, neural responses evoked by speech production spanned a broader set of cortical regions than those elicited during auditory perception.
Second, for channels that were highly responsive to both speech production and perception, the activity patterns across the two modalities were strongly correlated, with perception responses exhibiting a clear temporal delay relative to production.
Third, decoding performance was comparable between the left and right hemispheres across both speaking and listening tasks. 
In conclusion, our work (1) demonstrates the feasibility of a unified decoding framework applicable across both speaking and listening modalities, (2) exhibits hierarchical, character, and syllable generalization abilities, (3) employs a post-training and inference framework for LLMs to resolve the highly ambiguous one-to-many mapping from toneless syllables to logographic characters, improving sentence-level reconstruction, and (4) provides insights into the neural differences and similarities between Mandarin speech production and perception.

\section*{Results}
\label{results}
This section provides an overview of our decoding framework, data collection process, contribution analysis, and the decoding results. 
Detailed statistical procedures for all experimental analyses are provided in the Methods.

\subsection*{Brain-to-sentence decoding framework}
\label{brain-to-sentence-decoding-framework}
Our brain-to-sentence decoding framework consists of three components: a brain decoder that classifies the initials and finals of each Chinese character; a beam search module that generates multiple possible toneless syllable sequence candidates from the logits (i.e., output probabilities) produced by the brain decoder; and an LLM-based syllable-to-sentence decoder that derives the final sentence from the selected candidates.
Specifically, the brain decoder is composed of two identical modules, each trained in a supervised manner to discriminate initials and finals from the input brain signals. 
We adopted \nsk\cite{zhang2025neurosketcheffectiveframeworkneural}, an effective 2D-CNN neural decoder developed in our concurrent work, as the default classifier for initials and finals from neural recordings.
Then, for each sentence, we arranged the logits of the initials and finals for every character as predicted by the brain decoder, and performed beam search to retrieve the most probable decoding paths, each corresponding to a syllable sequence candidate. All candidates were ranked in descending order of their scores, and the top-20 candidates were finally retained.
Finally, we post-trained an LLM to generate the Chinese sentence from the  top-20 selected candidates. 
As shown in Fig.~\ref{fig:main_fig}e, before training, we expanded the LLM’s vocabulary to cover all toneless syllables in Mandarin Chinese. The model was then subjected to a three-stage supervised fine-tuning process, including a translation task that translated a syllable sequence into the corresponding Chinese sentence, a listwise ranking task that selected three candidates closest to the correct syllable sequence from the top-20 candidates, and a correction task that generated the correct sentence based on the three best candidates.
During inference, we employed a two-step decoding procedure (shown in Fig.~\ref{fig:main_fig}f): the top-20 candidates were fed into the post-trained LLM to select the top three, which were then input into the model again to generate the correct sentence.

\subsection*{Data collection}
\label{data-collection}
For data collection, intracranial recordings were obtained from 12 patients (denoted as S1 to S12, see Supplementary Table 1 for participant information) with drug-resistant epilepsy (7 males, 5 females; age range: 13–56 years). 
Electrode implantation covered both hemispheres in 6 patients, the left hemisphere in 3, and the right hemisphere in 3. Each patient was implanted with 7–17 electrodes, resulting in 107–205 bipolar-referenced channels per subject.
Our experimental paradigm was designed with two key considerations. First, all participants were patients with drug-resistant epilepsy, which limits the feasible duration for data collection. To obtain an adequate amount of data and ensure a training dataset balanced across syllables within a short time, we divided the data acquisition procedure for each participant into two parts (Fig.~\ref{fig:main_fig}a,b). 
The first part consisted of a single-character listening and speaking task, which served as the source of training data. 
The second part comprised a sentence-level listening and speaking task, which was used exclusively for evaluation.
Second, neural signals drift over time~\cite{Saha2018Evidence,Christensen2012EffectsVariability}, making it challenging to directly compare brain activity elicited by listening and speaking. 
To minimize temporal confounds, we interleaved the listening and speaking tasks, such that each character or sentence was spoken a few seconds after being heard, ensuring close temporal alignment between the two conditions and allowing for a more reliable comparison of the corresponding brain responses.

We constructed two distinct corpora (Supplementary 4), each containing separate training and test sets. The first corpus consisted of a training set comprising 49 Chinese characters, which included 11 initials, 15 finals, 22 toneless syllables, and 44 tonal syllables. The test set consisted of 22 sentences, each ranging from 4 to 11 characters in length, totaling 61 unique characters. Among these, 14 characters overlapped with the training set, and all syllables in the test sentences were derived from the 22 toneless syllables present in the training set. 
The second corpus featured a training set of 161 characters, covering 11 initials, 15 finals, 60 toneless syllables, and 159 tonal syllables. The test set consisted of 31 sentences, each ranging from 6 to 16 characters, containing 128 unique characters, including 66 toneless syllables. In the test set, 25 characters overlapped with the training set, while 17 toneless syllables were not present in the training set. 
Participants S1 to S10 used the first corpus; participants S11 and S12 used the second corpus.
Specifically, in the first corpus, each toneless syllable in the training set was presented between 20 and 60 times, whereas in the second corpus, each toneless syllable was repeated 15 times. Additionally, each test sentence in both corpora was presented twice.

\begin{figure}[tbp]
   \centering
   \includegraphics[width=1.0\textwidth]{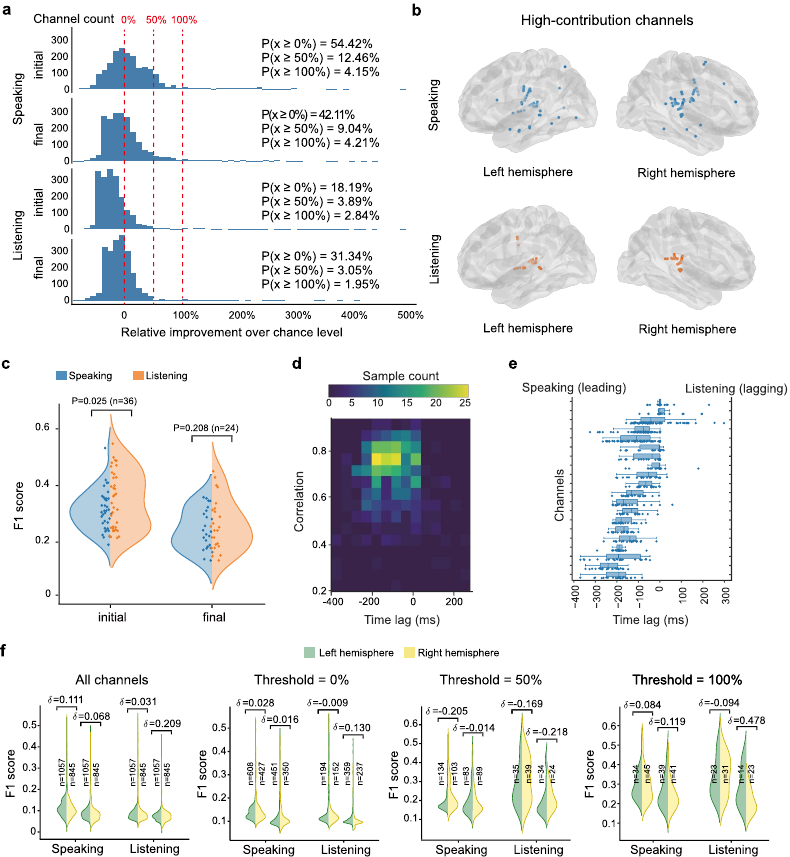}
   \caption{
   \textbf{Channel contribution analysis.}
   \textbf{a,} Histogram showing the distribution of relative improvement over chance level for all channels in the speaking and listening tasks. Three improvement thresholds, 0\%, 50\%, and 100\%, are indicated, corresponding to low-, medium-, and high-contribution channels, respectively.
   \textbf{b,} Spatial distributions of high-contribution channels identified during speech production and perception.
   \textbf{c,} Violin plots comparing the decoding performance between speaking and listening tasks for channels that were highly responsive to both speech production and perception.
   \textbf{d,} For the subset of channels in (c) exhibiting stable temporal relationships between modalities, a heatmap illustrating the maximum correlation between speaking- and listening-evoked neural signals and the corresponding time lags.
   \textbf{e,} Box plots showing the distribution of time lags across different syllables for each channel with stable speaking–listening delays.
   \textbf{f,} Violin plots comparing decoding performance between left and right hemispheric channels across different contribution thresholds.
   } 
   \label{fig:Fig2}
\end{figure}

\subsection*{Contribution analysis}
\label{contribution-analysis}
Across the 12 subjects in our dataset, a total of 146 depth electrodes were implanted, yielding 1902 channels after bipolar re-referencing, with 1057 located in the left hemisphere and 845 in the right. We first investigated the decoding contributions of these channels for speaking and listening to identify those with significant contributions. 
Only each participant’s training data was used and split into training and validation sets at a 4:1 ratio. Each channel was individually used to classify initials and finals with our brain decoder during both speaking and listening tasks, and the best validation performance was recorded.

\vpara{Channel contribution.}
In Extended Data Fig.~\ref{overall-spatial-distribution-of-channel-contributions}, we mapped the channel coordinates of all subjects onto a standard brain template and visualized the F1 scores of the channels in the classification tasks as heatmaps.
Meanwhile, we estimated the chance level of the classification task by randomly sampling labels according to the class distribution as weights and repeating the process 5000 times. 
Then, we computed the relative improvement of each channel’s performance over the chance level to quantify its contribution. In Fig.~\ref{fig:Fig2}a, we highlighted the channels corresponding to 0\%, 50\%, and 100\% improvement over the chance level, which represent low, medium, and high contributions, respectively.
The results show that, across all contribution levels, the proportion of channels in the speaking task was consistently higher than that in the listening task, indicating that brain activity associated with speech production is overall more widespread.
Fig.~\ref{fig:Fig2}b illustrates the spatial distribution of high-contribution channels during the speaking and listening tasks. Channels highly responsive to speech production were found across broader cortical regions, demonstrating that speech production engages a wider set of brain areas.
\vpara{Comparison between speech production and perception.}
After characterizing the spatial distribution of neural responses associated with speech production and perception, we next focused on channels that were responsive to both modalities.
To this end, we identified a subset of channels ($n=38$) exhibiting high contributions in both speaking and listening tasks, which were primarily distributed across the superior temporal and insular regions, and compared their neural response patterns during speech production and perception.
Fig.~\ref{fig:Fig2}c compares the decoding performance of these channels in the speaking and listening tasks. In the initial decoding, listening achieved higher overall accuracy than speaking ($p=0.025$), whereas in the final decoding, the performance of the two tasks was comparable ($p=0.21$). 

Motivated by these results, we further investigated the similarities and differences in brain activity evoked by speech production and perception.
Inspired by the analysis in \citet{Hamilton2021ParallelDistributed}, for each channel, we computed the maximum correlations and corresponding time lags between responses to the same syllables in speaking and listening conditions. We retained channels with stable time delays across samples (standard deviation less than 100 ms), resulting in 20 channels.
Fig.~\ref{fig:Fig2}d shows the relationship between the correlations and corresponding time lags, in which the correlations of these channels were notably high (mean = 0.7175, 90\% CI [0.4828, 0.8678]), indicating similar neural response patterns to identical syllables across speaking and listening. A similar phenomenon was revealed in the context of English speech by \citet{Chen2024CorticalRepresentation}. 
Additionally, the neural signals revealed a consistent latency in the listening condition relative to speaking. Fig.~\ref{fig:Fig2}e illustrates the distribution of time lags for each channel, showing consistent response latencies during listening compared to speaking (mean = –106.5 ms, 90\% CI [–249.4, 23.05]). 
This delay is consistent with established neural processing mechanisms of self-generated speech and perception of others~\cite{Magrassi2015SoundRepresentation}.
Collectively, these results suggest that channels responsive to both modalities exhibit similar response patterns with a clear temporal lag in speech perception relative to speech production.

\vpara{Comparison between left and right hemispheres.}
The left hemisphere of the brain is generally considered the dominant hemisphere for language and speech processing~\cite{Hickok2007CorticalOrganization}, and thus many studies on speech decoding have utilized neural signals recorded from the left hemisphere~\cite{Willett2023HighPerformance,metzger2023high,Card2024AccurateRapid}. \citet{Chen2024NeuralSpeech} found in their experiments that decoding of speech production could be accomplished using signals from either hemisphere, with no significant difference observed between the two.
Taking this finding a step further, we performed comparisons between the left and right hemispheres for both speech production and perception based on progressively more selective channel subsets. Starting from all channels, we subsequently examined three additional subsets exceeding the low-, medium-, and high-contribution thresholds, respectively. For each (sub)set, we grouped channels by hemisphere and compared the initial and final decoding performance between the left and right hemispheres.
As shown in Fig.~\ref{fig:Fig2}f, we compared the performance distributions of the left and right hemispheres and calculated their Cliff’s delta ($\delta$) effect sizes~\cite{Cliff1993DominanceStatistics}.
Across both the speaking and listening tasks, we did not observe any substantial differences between the left and right hemispheres. First, under all thresholds, the number of selected channels did not differ significantly between the two hemispheres. Moreover, among the 16 pairwise comparisons, 11 exhibited a Cliff’s delta with an absolute value below 0.147, indicating a small effect size~\cite{wan2019how}. Therefore, both the left and right hemispheres are potential targets for electrode implantation in the neural decoding of speech production and perception.

\begin{figure}[h]
   \centering
   \includegraphics[width=1.0\textwidth]{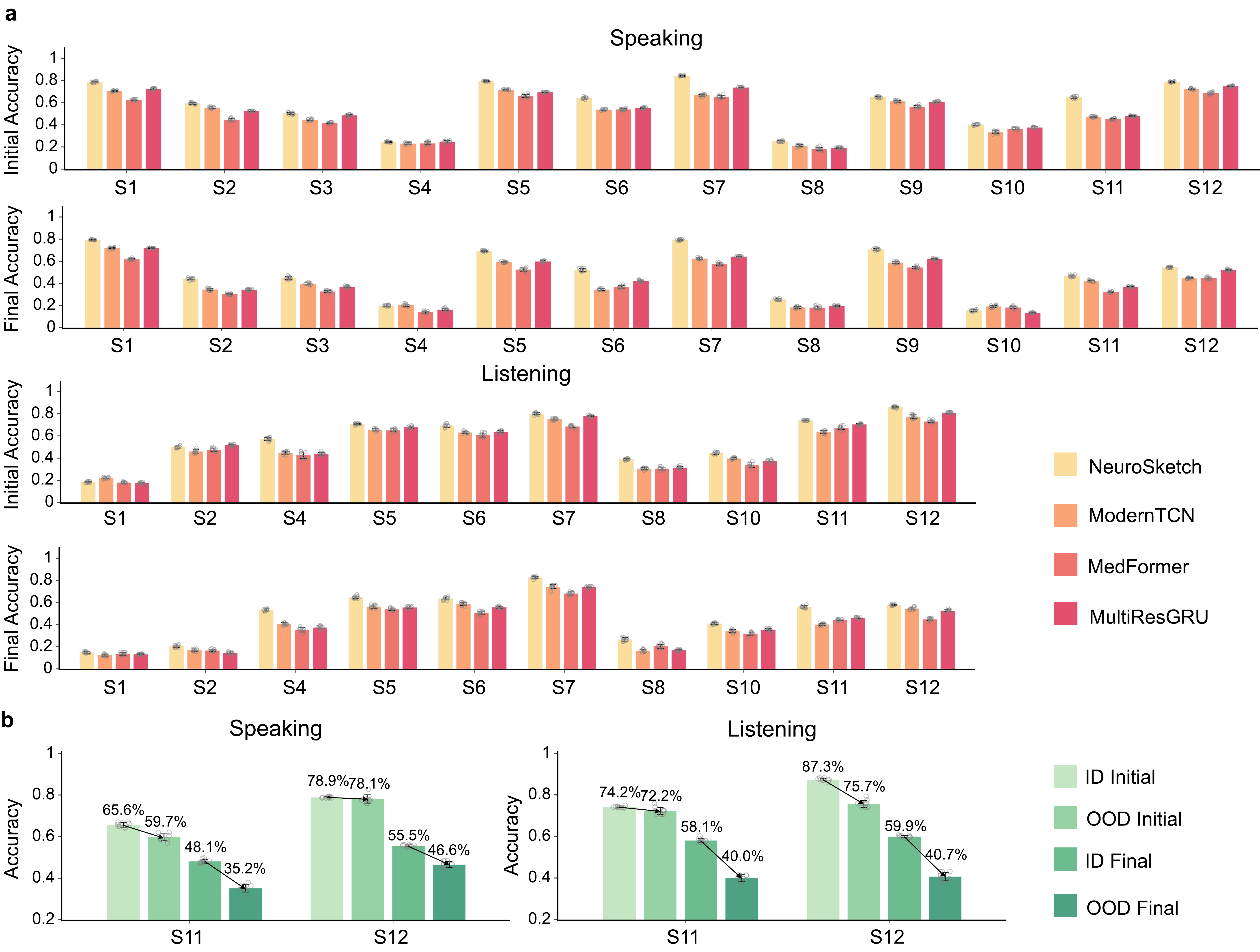}
   \caption{
   \textbf{Decoding of initials and finals.}
   \textbf{a,} Bar plots showing the initial and final classification accuracies of \nsk and three other brain decoders in the speaking and listening tasks. For each task, results are shown for participants with available high-contribution channels (12 out of 12 for speaking, 10 out of 12 for listening). Complete results for all participants are provided in the Supplementary Tables.
   \textbf{b,} Bar plots illustrating syllable generalization performance of \nsk. Classification accuracies of initials and finals are compared between in-domain (ID) and out-of-domain (OOD) syllables for participants evaluated on the second corpus.
   } 
   \label{fig:Fig3}
\end{figure}
\subsection*{Decoding of initials and finals}
\label{initial-final-classification}

Based on the results of contribution analysis, we conducted decoding experiments using high-contribution channels, i.e., channels whose activity patterns showed strong correlation with the decoding targets. Due to variability in electrode implantation locations and data quality across participants, two subjects (S3 and S9) did not have any high-contribution channels in the listening task. To focus on reliable decoding, we conducted our analysis on participants with available high-contribution channels in the corresponding tasks. Complete results for all participants are provided in the Supplementary Tables.

\vpara{Results overview.}
In the speaking task, we achieved average initial and final accuracies of 59.54\% (95\% CI [24.23\%, 84.10\%], representing a relative improvement of 394.9\% over chance level) and 50.17\% (95\% CI [14.94\%, 79.87\%], representing a relative improvement of 412.0\% over chance level), respectively, with the highest accuracies reaching 85.06\% for initials and 80.19\% for finals.
In the listening task, we achieved mean initial and final accuracies of 58.92\% (95\% CI [17.69\%, 86.17\%], representing a relative improvement of 389.7\% over chance level) and 48.05\% (95\% CI [14.29\%, 82.96\%], representing a relative improvement of 406.6\% over chance level), respectively, with top accuracies of 86.76\% and 83.77\%, respectively.
At the individual-participant level, substantial dissociations between speaking and listening performance were observed, most commonly manifesting as markedly higher decoding accuracy during speech production than during perception.
For example, participant S1 achieved initial and final decoding accuracies of 78.64\% (95\% CI [77.60\%, 79.55\%]) and 79.38\% (95\% CI [78.97\%, 79.87\%]) in the speaking task, whereas the corresponding accuracies dropped to 18.35\% (95\% CI [17.53\%, 19.16\%]) and 14.87\% (95\% CI [14.03\%, 15.91\%]) in the listening task. Similarly, participant S9 showed high decoding performance during speaking (mean = 64.93\%, 95\% CI [63.96\%, 65.84\%] for initials and mean = 70.81\%, 95\% CI [69.45\%, 71.75\%] for finals) but did not exhibit any high-contribution channels in the listening task, resulting in initial decoding accuracy close to chance level during perception ($p=0.92$).
For some participants, decoding performance in the listening task exceeded that in the speaking task. However, the resulting performance gaps were smaller than those observed when speaking markedly outperformed listening.

\vpara{Comparison of different brain decoders.}
We compared \nsk (2D CNN–based) with three other architectures, encompassing mainstream deep neural network paradigms, including \tcn\cite{donghao2024moderntcn} (1D CNN–based), \medformer\cite{wang2024medformer} (Transformer–based), and \gru\cite{zinxira2024tlvmc}(RNN–based).
The bar plots in Fig.~\ref{fig:Fig3}a and Extended Data Fig.~\ref{f1-scores-of-initial-final-classification} summarize the classification performance of the four brain decoders. 
Across the four brain decoders, \nsk achieved the highest overall accuracy. Specifically, in the speaking task, \nsk attained average accuracy improvements of 6.44\% and 7.66\% (both $p<0.001$) over the second-best model (\gru) in initial and final classification, respectively. In the listening task, \nsk showed 4.59\% and 7.61\% higher average accuracies than the second-best models, \gru for initials and \tcn for finals, respectively (both $p<0.001$).
\gru demonstrated a modest overall advantage over \tcn on initial classification, achieving 1.32\% and 1.51\% higher accuracies in the speaking and listening tasks (both $p<0.001$).
In contrast, the two models showed comparable performance on final classification ($p=0.47$ in the speaking task and $p=0.08$ in the listening task). 
Among the four brain decoders, \medformer showed relatively lower performance compared to the other architectures, with the difference being statistically significant ($p < 0.001$) across both tasks.

\vpara{Out-of-domain syllable generalization.}
\label{out-of-domain-syllable-generalization}
A practical challenge for neural speech decoding in Mandarin Chinese is the large size of its syllable inventory, which includes over 400 toneless syllables and more than 1200 tonal syllables.
Achieving reliable decoding for the full syllable inventory requires extensive data collection.
For instance, \citet{qian2025real} reported a 13-day recording protocol for a single participant, with 2–3 hours of recording per day, during which the participant produced 394 toneless syllables, each repeated approximately 30 times.
In contrast, our data collection typically lasted only 2–3 hours for most participants within a single day, which made it impossible to sample a large number of syllables extensively.
This constraint highlights the importance of decoding syllables absent from the training set, enabling the model to extend from a limited subset of observed syllables to the broader syllabic space.

We evaluated the out-of-domain (OOD) syllable generalization performance of \nsk by dividing the test set of the second corpus into in-domain (ID) and OOD syllables.
In OOD syllables, we achieved average accuracies of 71.42\% (95\% CI [58.58\%, 79.40\%]) for initials and 40.60\% (95\% CI [34.40\%, 48.28\%]) for finals, representing relative improvements over their respective chance levels of 571.2\% and 333.3\%, respectively. The results demonstrate that our framework can generalize to OOD syllables and accurately classify their initials and finals.
Fig.~\ref{fig:Fig3}b compares the accuracies of the initials and finals on the two subsets. The average accuracy for initials on OOD syllables decreased by 5.08\% compared to that on ID syllables, while finals exhibited a larger drop of 14.80\%.
This phenomenon suggests that finals were more difficult to decode when generalizing to previously unseen syllables.
One possible explanation is that finals exhibit greater acoustic and articulatory variability across syllables. For example, the duration of the final \textit{a} differs between the syllables \textit{pa} (shorter and more abrupt) and \textit{ma} (typically longer and more nasalized). In addition, the articulatory configuration of the final \textit{i} varies with the preceding initial. For instance, the \textit{i} in \textit{shi} is a retroflex vowel produced with the tongue curled back, whereas in \textit{ji} it corresponds to a high front vowel.
As a preliminary exploration, these results demonstrate that an initial–final–based decoding paradigm can achieve OOD syllable generalization, providing empirical evidence that Mandarin syllable decoding does not necessarily require full coverage of the syllabic inventory.

\subsection*{Sentence decoding}
\label{sentence-decoding}
\vpara{Quality of syllable sequence candidates.}
Next, we analyzed the syllable sequence candidates generated from beam search, which was performed on the predicted probabilities output by the brain decoder. 
To assess the quality of the candidates from beam search, we utilized the syllable error rate (SER) distribution of the top-20 candidates and the exact match probability (EMP), which is defined as the probability that the top-20 candidates contain perfectly correct syllable sequences.

\begin{figure}[H]
   \centering
   \includegraphics[width=1.0\textwidth]{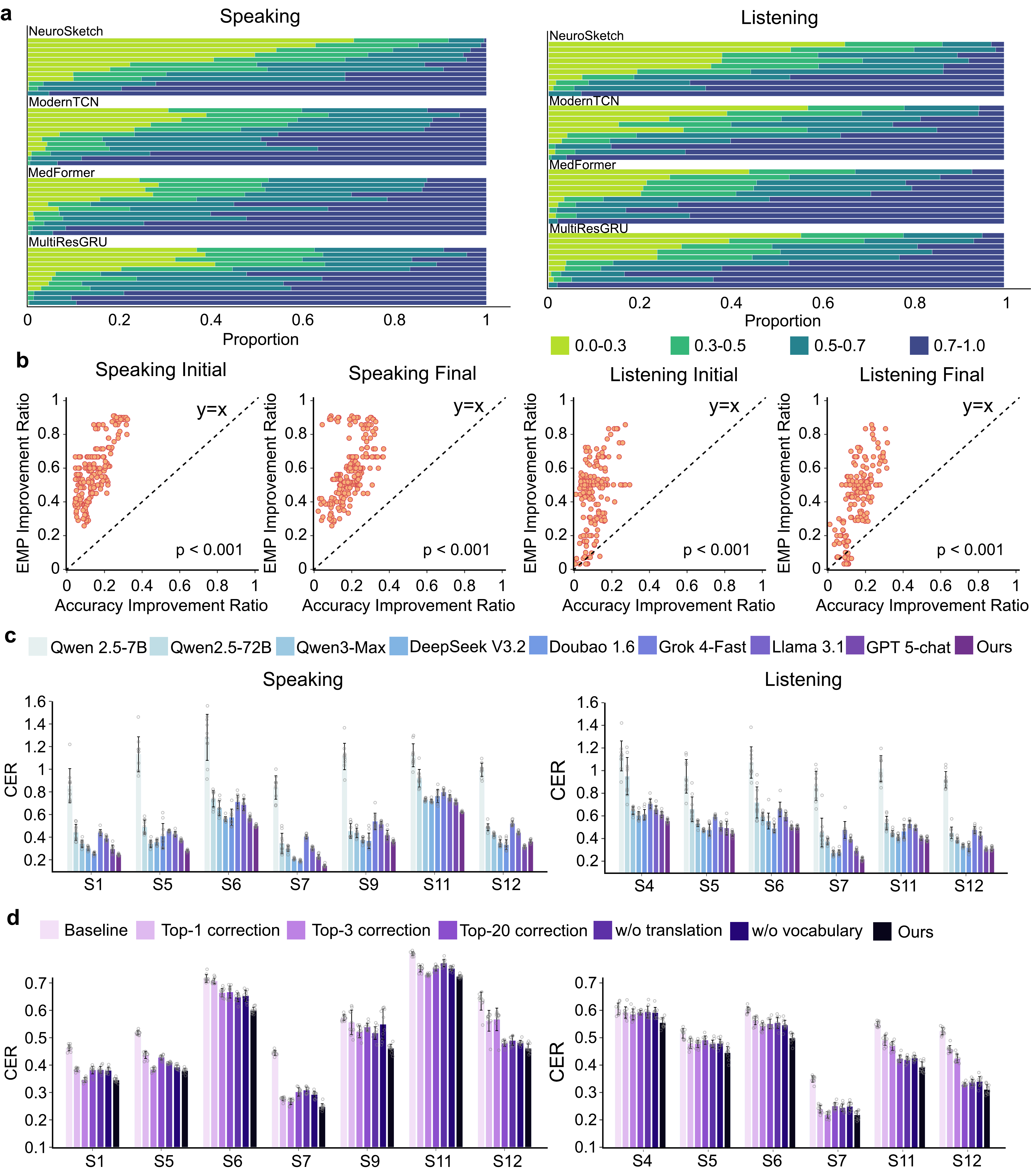}
   \caption{
   \textbf{Sentence decoding.}
   \textbf{a,} Proportion distributions of SER for the top-20 syllable-sequence candidates generated by \nsk and the three other brain decoder architectures in the speaking (12 participants) and listening (10 participants) tasks.
   Complete results of our model for all participants are provided in the Supplementary Tables.
   \textbf{b,} Scatter plot illustrating the relationship between the relative improvement in initial and final classification accuracies and the relative improvement in EMP among the top-20 candidates.
   \textbf{c,d,} Results for sentence-level decoding. To ensure that the input candidate sets contain a minimal threshold of reliable information for sentence decoding, we focused on participants whose average proportion of high-quality candidates (SER $< 0.3$) across the decoding corpus exceeded 10\% (7 out of 12 for speaking, 6 out of 12 for listening).
   Complete results of our model for all participants are provided in the Supplementary Tables.
   \textbf{c,} Bar plots comparing CER of our LLM and other LLMs on the syllable-to-sentence decoding task, including small-sized LLM (\llmbase), medium-sized LLMs (\qwenmiddle), and large-scale commercial LLMs (\qwenmax, \deepseek, \doubao, \gpt, \grok, \llama).
   \textbf{d,} Bar plots presenting ablation results of our syllable-to-sentence decoding framework. The Baseline method used an IME-style decoder based on lexicon- and language-model–driven search. The Top-1, Top-3, and Top-20 correction groups performed direct correction of the top-k beam search candidates. Auxiliary component ablations, labeled ``w/o translation'' and ``w/o vocabulary,'' were created by removing the respective components from the full framework.
   } 
   \label{fig:Fig4}
\end{figure}

Fig.~\ref{fig:Fig4}a shows the SER distribution of the beam search results. 
In the speaking task, across all subjects, the proportion of high-quality candidates (SER $<$ 0.3) and the EMP produced by \nsk were 28.23\% and 27.65\%, respectively. These values are approximately twice those of the second-best model (\gru), which achieved 15.66\% high-quality candidates and an EMP of 12.86\%. 
Furthermore, we investigated the impact of changes in the initial-final classification accuracy on the EMP. Using the performance of \nsk as a reference point, we calculated the improvement ratios in initial-final classification accuracy and EMP compared to other models. 
As shown in Fig.~\ref{fig:Fig4}b, the improvement ratio in the EMP significantly exceeded that of the initial-final classification accuracy ($p < 0.001$). When the classification accuracy improved by approximately 20\%, the corresponding EMP could increase by over 80\% for some samples, indicating a substantial enhancement in candidate quality.
The above results demonstrate that the advantage in initial-final classification was amplified during the beam search stage, indicating that a well-performing brain decoder is crucial for decoding performance.

\vpara{Syllable-to-sentence decoding using LLM.}
The final step of the decoding framework leverages an LLM to generate the spoken or heard sentence from multiple toneless syllable sequence candidates. To tackle this challenging problem, we decomposed it into two subtasks.
As shown in the preceding analyses, errors in initial and final decoding were amplified during sequence-level decoding. Consequently, when initial and final decoding accuracy is insufficiently high, the beam-search candidate sets can deviate substantially from the correct sequences.
In most such cases, all candidates within the beam exhibited very high SERs, making it impossible to reconstruct the original sentences and thereby hindering effective evaluation of the sentence decoding performance. 
Therefore, we focused our analysis on participants whose average proportion of high-quality candidates (SER $< 0.3$) across the decoding corpus exceeded 10\%, which served as a minimal quality threshold below which reliable sentence reconstruction was rarely possible. Complete sentence decoding results of our model for all participants are available in the Supplementary Tables.

Fig.~\ref{fig:Fig4}c presents the sentence-level decoding performance of these participants.
In the speaking task, we achieved an average CER of 35.92\% across 7 participants (95\% CI [14.19\%, 62.58\%]), and in the listening task, the average CER across 6 participants was 40.31\% (95\% CI [20.28\%, 57.06\%]).
Among these participants, 5 achieved reliable sentence decoding in both the speaking and listening tasks, with average CERs of 38.18\% (95\% CI [14.03\%, 62.64\%]) and 37.28\% (95\% CI [20.12\%, 51.68\%]), respectively.
In the first corpus, we achieved the best CERs of 14.71\% (95\% CI [13.27\%, 16.97\%]) and 21.80\% (95\% CI [19.56\%, 23.58\%]) for the speaking and listening tasks. In the second corpus, where the test sentences included syllables not present in the training set, the best CERs of the speaking and listening tasks were 36.06\% (95\% CI [33.83\%, 38.51\%])and 31.10\% (95\% CI [28.98\%, 33.61\%]), respectively.
In the following results, we tested the performance of other LLMs as well as variants of our own model. Given that neural decoding typically requires rapid response times, we disabled the deep-thinking mode during inference for all models, using standard inference exclusively.

As our model was based on \llmbase~\cite{qwen2025qwen25technicalreport}, we first tested the performance of the original \llmbase on the syllable-to-sentence task. It was observed that the average CERs of \llmbase were 104.84\% (95\% CI [70.25\%, 146.25\%]) and 99.21\% (95\% CI [71.66\%, 129.44\%]) in the speaking and listening tasks, respectively, indicating that directly applying a 7B model to this task is entirely infeasible.
For medium-sized models, we tested \qwenmiddle and found a significant improvement over \llmbase ($p < 0.001$), with the average CER in the speaking task (mean = 55.38\%, 95\% CI [26.58\%, 95.77\%]) reduced to approximately half that of \llmbase.
Furthermore, we evaluated six large-scale commercial LLMs, including the Chinese-oriented LLMs \qwenmax~\cite{yang2025qwen3}, \deepseek~\cite{liu2025deepseek}, and \doubao~\cite{doubao16}, as well as the English-oriented LLMs \gpt~\cite{gpt5chatlatest}, \grok~\cite{grok4fast}, and \llama~\cite{grattafiori2024llama3herdmodels}.
The results show that most of these large-scale LLMs significantly outperformed small- or medium-sized LLMs ($p < 0.001$). 
Among the evaluated LLMs, \gpt, \doubao, and \deepseek showed comparable leading performance, with pairwise significance tests between these models yielding $p$-values greater than 0.05 in both the speaking and listening tasks.
Even though the task primarily involves understanding Hanyu Pinyin, Chinese-oriented LLMs did not demonstrate a significant performance advantage over English-oriented LLMs. 
In addition, we compared our model with these commercial LLMs. 
Across both speaking and listening tasks, our model achieved significantly lower average CERs than all commercial LLMs ($p < 0.001$). Specifically, it outperformed the best commercial models by an average margin of 4.97\% in the speaking task (compared with \deepseek) and 3.10\% in the listening task (compared with \gpt).
These results demonstrate that decomposing the task into two subtasks and applying appropriate post-training can substantially enhance LLM performance.

We also conducted an ablation study to evaluate the effectiveness of the proposed LLM-based syllable-to-sentence decoding framework, examining key design choices related to LLM post-training, beam-search candidate retention, explicit task decomposition, and auxiliary components.
Fig.~\ref{fig:Fig4}d shows the sentence decoding results of our model and ablation groups.
We began by constructing a minimal baseline in which only the top-1 beam-search syllable sequence was retained and converted into a Chinese sentence using an input method editor (IME)-style decoder based on lexicon- and language-model–driven search, representing a conventional syllable-to-text decoding strategy.
Under this setting, the baseline achieved an average CER of 49.31\% (95\% CI [33.97\%, 71.13\%]) in the speaking task and 52.42\% (95\% CI [34.30\%, 62.90\%]) in the listening task.
We then replaced the IME-style decoder with a post-trained correction LLM while keeping all other components unchanged.
This modification led to an average CER reduction of 6.86\% ($p<0.001$) in the speaking task and 5.38\% ($p<0.001$) in the listening task, indicating that LLM-based correction exhibits greater potential to resolve syllable-to-sentence decoding than conventional lexicon- and language-model–based decoding.
Next, we examined the effect of retaining multiple beam-search candidates. When the number of candidates was increased from one to three, decoding performance improved by 2.80\% ($p<0.001$) and 1.76\% ($p<0.001$) in the speaking and listening tasks, respectively.
However, further increasing the candidate set to 20 did not lead to consistent additional improvements, with performance remaining comparable to that of the top-3 setting ($p=0.27$ in the listening task).
This saturation effect indicates that, under direct correction, expanding the candidate set increases the proportion of low-quality hypotheses, which can obscure the correct solution and limit further performance gains.
Notably, in our framework, when direct correction was decomposed by first selecting the most promising candidates from the larger set using a listwise ranking strategy before applying correction, performance continued to improve with statistical significance ($p<0.001$), underscoring the importance of structured task decomposition for effectively leveraging large and noisy candidate sets.
Finally, we evaluated two auxiliary components of the framework: extending the LLM vocabulary to fully cover toneless syllables and performing a preliminary translation post-training step.
Removing either component resulted in a performance degradation ($p<0.001$) in both the speaking and listening tasks, confirming that these design choices are integral to the overall decoding pipeline.
In conclusion, these ablation results demonstrate that the proposed syllable-to-sentence decoding framework benefits from both high-level framework design choices and fine-grained component-level optimizations.

\begin{figure}[htbp]
   \centering
   \includegraphics[width=1.0\textwidth]{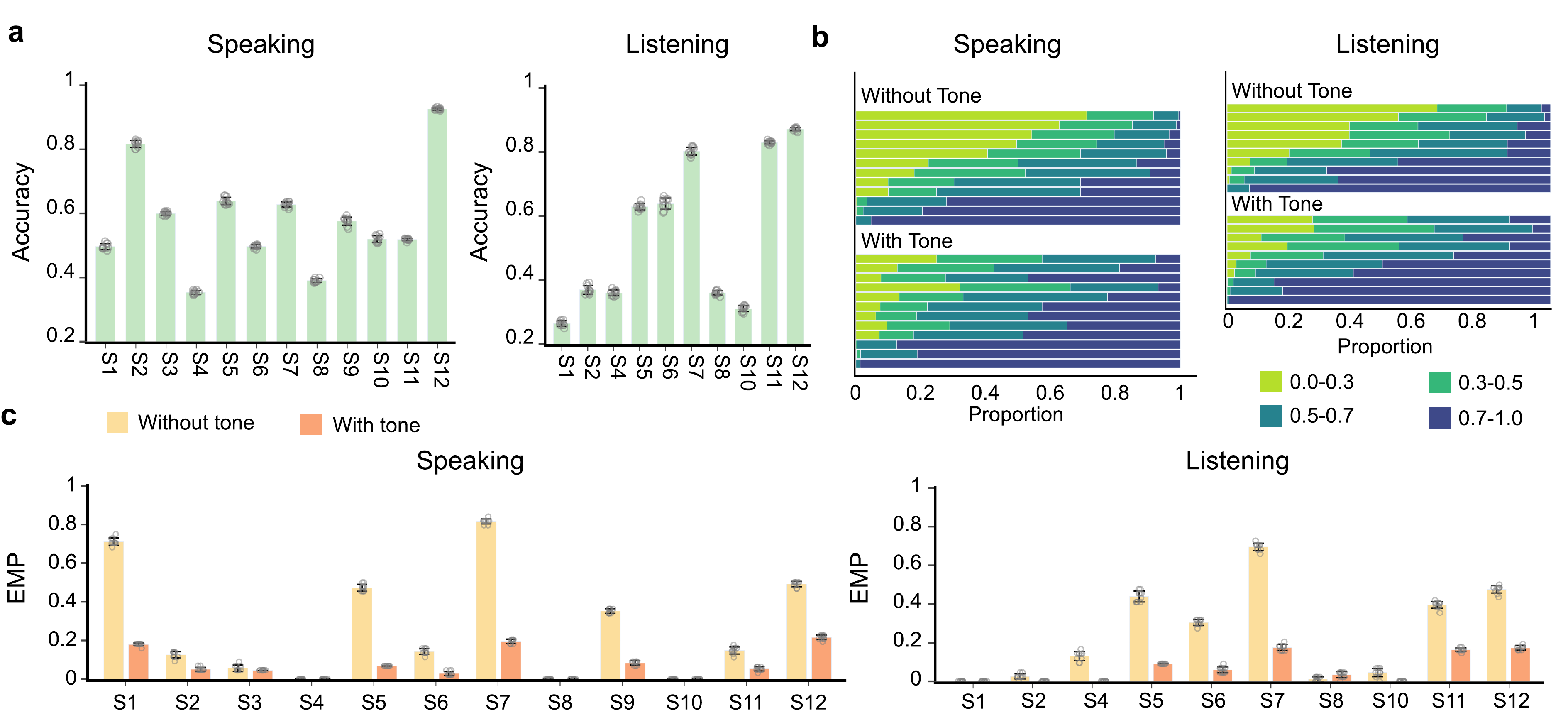}
   \caption{
   \textbf{Decoding of tones.}
   \textbf{a,} Bar plots showing tone classification accuracy in the speaking and listening tasks.
   \textbf{b,} Proportion distributions of syllable error rates for the top-20 syllable sequence candidates with and without tone information.
   \textbf{c,} Bar plots comparing exact match probability under decoding with and without tone information.
   } 
   \label{fig:Fig5}
\end{figure}
\subsection*{Decoding of tones}
\label{decoding-of-tones}
Hanyu Pinyin is composed of three fundamental elements: initials, finals, and tones. Previous results demonstrated that sentence-level decoding could be achieved using only initials and finals, combined with the ability of an LLM. 
In this section, we explain why tone information is not decoded in our framework. To investigate this, we trained our brain decoder to classify tones and applied beam search to generate tone-bearing syllable candidates using the predicted probability distributions over initials, finals, and tones.

Fig.~\ref{fig:Fig5}a and Extended Data Fig.~\ref{f1-scores-of-tone-classification} report the results of tone classification. Overall, the average tone classification accuracy was 58.01\% (95\% CI [35.14\%, 92.50\%]) in the speaking task and 54.36\% (95\% CI [25.97\%, 87.35\%]) in the listening task. We found that, although tone has only four categories, the classification accuracies in many subjects (4 subjects in the speaking task and 4 subjects in the listening task) were lower than those of initials and finals, which have more than ten categories, indicating that the neural signal patterns for tones are less distinct than those for initials and finals. 
Meanwhile, the relatively high error rate in tone classification is further amplified during beam search. Fig.~\ref{fig:Fig5}b,c compare the quality of the beam search candidates generated without and with tone information.
In the speaking task, incorporating tones substantially degraded candidate quality: the average proportion of high-quality candidates dropped from 28.23\% to 10.09\%, falling to approximately one third of the original level.
A similar trend was observed in the listening task, where the high-quality proportion decreased from 25.94\% to 9.74\%.
Consistently, the average EMP declined sharply when tones were included: in the speaking task from 27.65\% to 7.69\%, and in the listening task from 25.23\% to 6.95\%, both falling to less than one-third of the original performance. 
These results demonstrate that introducing tones drastically reduces the quality of the decoded syllable-sequence candidates.
Moreover, our sentence decoding experiments showed that both current commercial LLMs and our model can successfully infer correct sentences from toneless syllable sequences, demonstrating that tone information is not necessary for sentence decoding.
Therefore, tones were not incorporated in our decoding framework.

\section*{Discussion}
\label{discussion}
In this work, we introduce a unified approach to brain-to-sentence decoding for both speaking and listening in Mandarin Chinese. 
By integrating the two modalities within a unified experimental and modeling framework and leveraging the broad spatial coverage of sEEG electrodes, we successfully decoded full sentences in both modalities, demonstrating the potential of multimodal brain–language decoding.
A notable aspect of our decoding framework is its generalization ability. Even when trained only on single-character speaking or listening data, the framework can decode full sentences. It also handles characters and syllables that never appeared during training, demonstrating broad generalization across linguistic units.
To address the high ambiguity inherent in mapping Mandarin syllables to characters and words, we propose an LLM-based syllable-to-sentence decoding framework that performs end-to-end mapping directly from a set of syllable sequence candidates to the corresponding Chinese sentence.
Given the difficulty of this task for LLMs, we introduce a principled approach leveraging task decomposition alongside continued post-training of the LLM. Specifically, we develop a three-stage post-training and two-stage inference framework, which enables a 7-billion-parameter LLM to outperform several much larger commercial models in this challenging decoding task.

Using this unified framework, we not only achieved sentence decoding but also enabled direct comparison between the two modalities. 
Our key observations include the following: First, neural responses during speech production were distributed across broader cortical regions than those observed during speech perception. Second, channels jointly responsive to speech production and perception showed highly similar neural dynamics across the two modalities, with perceptual responses consistently lagging behind those observed during production. Third, decoding performance was comparable between the left and right hemispheres for both speaking and listening tasks.  
These results highlight shared and distinct neural characteristics of speech production and perception.

Despite the promising results achieved in this study, several limitations remain. 
First, our LLM-based syllable-to-sentence decoding framework requires the full sentence to be completed before generating outputs, which prevents online decoding of each character as it is being spoken or heard. 
Second, while our results demonstrate that initials maintain relatively high accuracy in OOD syllable generalization, finals exhibit a larger performance drop.
Third, although we achieved decoding performance that was consistently and substantially above chance level, the confidence intervals reveal pronounced inter-subject variability.
This variability primarily arises from inherent challenges in clinical neural recordings, including differences in electrode implantation sites, signal quality, available training samples, and participant engagement.
How to mitigate such data heterogeneity at the decoding level, particularly for subjects with less informative neural signals, and further extend brain-to-text decoding toward robust cross-subject generalization with sEEG remains an open and challenging problem that warrants future investigation.

Future work will focus on addressing these challenges. 
To improve online decoding, future work may explore frameworks that support incremental syllable-to-character conversion, allowing for continuous output updates based on partial input and historical decoding context. 
Regarding OOD syllable generalization, inspired by \citet{Feng2025AcousticInspired}, we will further explore more fine-grained acoustic feature-based segmentation of finals, explicitly distinguishing different realizations of the same final to guide the model toward learning more transferable representations.
With respect to cross-subject decoding, 
\citet{Singh2025TransferLearning} demonstrated successful reconstruction of English phoneme sequences using a model with a shared sub-component across participants, highlighting the potential of cross-subject brain-to-sentence decoding with sEEG.
To mitigate sEEG data heterogeneity and facilitate more robust and generalizable brain-to-text decoding, future work may explore complementary directions at both the data-analysis and model-design levels.
From a data perspective, we aim to investigate intrinsic properties of sEEG signals by identifying neural features that are shared across subjects under specific conditions, such as within homologous cortical regions. If such shared representations prove to be sufficiently generalizable and informative, they may enable training strategies that leverage pooled data across participants.
In parallel, at the model-design level, we plan to explicitly encourage the learning of these shared neural representations while suppressing subject-specific or task-irrelevant variations, thereby providing a potential pathway toward more robust and generalizable brain-to-text decoding.

Beyond the immediate technical contributions, our work carries broader implications across multiple domains. 
Within the field of artificial intelligence, although instantiated with Mandarin Pinyin initials and finals, this decoding paradigm can be broadly applied to any sensor-to-symbol pipeline capable of generating an n-best hypothesis set, thus opening avenues for versatile neural decoding across different modalities and languages.
Moreover, the syllable-to-sentence decoding framework we developed demonstrates the great potential of LLMs for neural language decoding. Unlike traditional approaches that treat LLMs merely as downstream error-correction tools applied to decoded sentences, our framework enables LLMs to directly process more diverse input forms and perform more complex tasks with greater flexibility and effectiveness.
In the context of brain-computer interfaces (BCI), these advances support progress toward neural decoding systems that handle multiple language input and output modalities.
From a neuroscience perspective, our observations offer insights into the shared and distinct neural processes underlying language production and perception, which may guide future investigations into the neural basis of multimodal language processing.

\section*{Methods}

\subsection*{Data acquisition}
\label{data-acquisition}

\vpara{Participants.}
Twelve participants with drug-resistant epilepsy were enrolled from three tertiary hospitals in China: Huashan Hospital (Fudan University), Chongqing Xinqiao Hospital, and the First Affiliated Hospital of Fujian Medical University. All underwent clinical sEEG monitoring as part of their pre-surgical evaluation. Participants were native Mandarin speakers, and written informed consent was obtained before data collection, in accordance with institutional ethics guidelines and the Declaration of Helsinki. The study was approved by the Huashan Hospital Institutional Review Board of Fudan University (HIRB, KY2019-518).

\vpara{Signal recording.}
The electrode implantation plan, encompassing the location and number of implants, was devised solely for the treatment of epilepsy. Neural signals were recorded using a multi-channel electrophysiological recording system (EEG-1200C, Nihon Kohden, Tokyo, Japan) at a sampling rate of 2000Hz. Concurrently, audio was acquired via a microphone placed in proximity to the participant to ensure clear and reliable sound recording, sampled at 44.1 kHz. Both neural and audio signals were synchronized and recorded in real time using the BCI2000 software platform (\url{https://www.bci2000.org}).

\vpara{Electrode anatomy localization.}
Individual brain reconstructions were performed using preoperative T1-weighted MRI with FreeSurfer~\cite{Fischl2012} (version 7.3.2). Postoperative CT was co-registered to the preoperative MRI using VeraView (United Imaging Healthcare), a clinical imaging platform that enables rigid fusion of multimodal volumes through intensity-based alignment. Intracranial electrode contacts were manually annotated on the coregistered CT–MRI overlay within VeraView, with spatial localization guided by anatomical landmarks. The resulting electrode coordinates in native space were subsequently used for surface-based mapping and further transformation into standard MNI space. Supplementary Fig.1 shows the electrode locations for each participant.



\vpara{Acoustic contamination.}
We conducted a contamination analysis to rule out the possibility that our neural signals are contaminated by acoustic noise, utilizing the method proposed in \citet{Roussel2020ObservationAcoustic}. 
Firstly, we calculated the correlations between frequency components in the audio and the neural signals to obtain the contamination matrix. The mean value on the diagonal
was computed to obtain a contamination index. 
Then, a distribution of surrogate indices was built by computing the mean diagonal 10000 times on as many shuffled versions of the contamination matrix.
Each time, we shuffled the contamination matrix by permuting either its rows or its columns.
Finally, we calculated the proportion $P$ of surrogate indices that exceeded the original index, where $P > 0.05$ indicates insufficient evidence to reject the null hypothesis of no contamination.
Supplementary Fig.2a presents an example contamination matrix and the corresponding distribution of surrogate indices for participant S6.
As shown in Supplementary Fig.2b, all participants exhibited $P$ values greater than 0.05, suggesting no evidence of acoustic contamination in our neural recordings.

\subsection*{Experimental paradigm}
\label{task-description}
\vpara{Data collection for training.}
Each participant was guided by visual cues to perform trial tasks. Each trial began with a white fixation cross presented centrally on a black background for 2 seconds, followed by a 1-second auditory cue, a 1-second inter-stimulus interval, and a 3.5-second articulation window. The articulation window began with a 1-second ready period, followed by a 2.5-second dynamic horizontal progress bar that guided sustained vocalization until its completion. Due to clinical constraints and variability in recording stability across patients, the experimental duration was limited to 2-3 hours, yielding 5-15 sessions for each participant. 

\vpara{Data collection for evaluation.}
For the first corpus, each experimental session lasted approximately 13 minutes, including one presentation of each sentence. For the second corpus, completing one full pass over all test sentences required two sessions.
Participants were guided by visual cues to perform trial tasks. Each trial began with a 2-second preparatory period, followed by an auditory prompt containing the full sentence. Within the audio prompt, each character was presented for 1 second, with 1 second inter-character intervals. 
Following a 1-second inter-stimulus interval, a dynamic horizontal progress bar appeared on the screen to guide sentence articulation. Each character within the sentence was allocated 3 seconds, which comprised a 1-second inter-character pause and 2 seconds of sustained vocalization.

\subsection*{Data preprocessing}
Neural signals were first visually inspected to exclude channels exhibiting excessive noise or absent activity. The remaining data were downsampled to 1000 Hz, band-pass filtered between 0.5 and 200 Hz, and notch filtered at 50 Hz to remove power-line interference. Bipolar re-referencing was then applied to enhance spatial specificity, and each channel was subsequently normalized via z-score transformation. 
We segmented neural signals based on characters. In the listening task, each character’s auditory cue lasted 1 second, so we used a uniform 1-second window for segmentation. In the speaking task, individual character articulation involved a 3.5-second window, beginning with a 1-second ready period followed by a 2.5-second progress bar that guided sustained vocalization. Accordingly, we segmented these trials using a 2.5-second window. For sentence-level speech, each character was allocated 3 seconds, consisting of a 1-second inter-character pause and 2 seconds of sustained vocalization. We segmented these with a 2-second window and applied zero-padding before and after to match the 2.5-second window length used for individual characters.

\subsection*{Response latency analysis}
\label{response-lantency-analysis}
Here we describe the implementation details of the response latency analysis.
In the analysis, we selected channels that were highly responsive to both speech production and perception, and compared their neural signals under the two modalities. 
We utilized the data in the training set, in which participants either listened to or produced individual characters. Neural signals here were aligned to the actual behavioral onset, defined as the audio playback onset in the listening task and the speech onset in the speaking task. For each trial, we extracted a 1-second neural segment starting from the corresponding onset. Signals associated with the same syllable were then averaged separately for the speaking and listening conditions.

To quantify the temporal similarity and relative delay between neural responses evoked by speech production and perception, we performed a cross-correlation analysis on the neural signals recorded from individual channels.
Given two single-channel neural signals $x(t)$ and $y(t)$, corresponding to the same syllable under speaking and listening conditions, respectively, we first removed their mean values to eliminate direct-current components:
\begin{equation}
\tilde{x}(t) = x(t) - \frac{1}{N}\sum_{t=1}^{N} x(t), \quad
\tilde{y}(t) = y(t) - \frac{1}{N}\sum_{t=1}^{N} y(t),
\end{equation}
where $N$ denotes the length of the two signals.
We then computed the cross-correlation function between the mean-centered signals as:
\begin{equation}
R_{xy}(\tau) = \sum_{t} \tilde{x}(t)\,\tilde{y}(t + \tau),
\end{equation}
where the lag $\tau$ spans the full range from $-(N-1)$ to $(N-1)$. To focus on physiologically plausible temporal offsets, the analysis was restricted to lags satisfying $|\tau| \leq \tau_{\max}$.
To facilitate comparison across channels and conditions, the cross-correlation function was normalized to yield a correlation coefficient:
\begin{equation}
\rho_{xy}(\tau) =
\frac{R_{xy}(\tau)}
{\sqrt{\sum_t \tilde{x}(t)^2 \sum_t \tilde{y}(t)^2}},
\end{equation}
which bounds the correlation values between $-1$ and $1$.
For each channel, we identified the maximum correlation coefficient
\begin{equation}
\rho_{\max} = \max_{\tau} \rho_{xy}(\tau),
\end{equation}
along with the corresponding lag $\tau^{*}$. The value $\rho_{\max}$ was used to quantify the similarity between speaking- and listening-evoked neural responses, while $\tau^{*}$ was interpreted as the relative temporal delay between the two conditions.

\subsection*{Brain decoder}
\label{brain-decoder}
In this section, we provide detailed descriptions of \nsk and three other brain decoders evaluated in our work. 
\nsk is described in a separate work evaluated on public datasets. The present study is independent in terms of data and experimental setting, and focuses on a unified multimodal brain-to-sentence decoding framework and LLM-based inference for Mandarin.
The schematic diagrams of the architectures for the four brain decoders are presented in Extended Data Fig.~\ref{architectures of brain decoders}. The input to each decoder can be defined as $\mathrm{X} \in \mathbb{R}^{C \times T}$, where $C$ denotes the number of channels and $L$ denotes the length of the neural recording acquired during speech production or perception. 

\vpara{\nsk.}
\nsk\cite{zhang2025neurosketcheffectiveframeworkneural} is a 2D-CNN-based neural decoding architecture developed in our concurrent work.
Given the input $\mathbf{X} \in \mathbb{R}^{C \times L}$, the model first reshapes it into $\mathbf{X'} \in \mathbb{R}^{B\times 1 \times 3C \times \lfloor L/3\rfloor}$. The reshaped two-dimensional representation is subsequently processed by a stem stage comprising four sequential stem blocks. Each block consists of a two-dimensional convolutional layer, followed by batch normalization and a ReLU activation function. The convolutional layers in all four blocks utilize a kernel size of 3, padding of 1, and strides of 2, 1, 1, and 2, respectively. The input and output feature dimensions for the four blocks are as follows: from 1 to 64, 64 to 32, 32 to 64, and 64 to 96, respectively.
Following the stem stage, the network consists of four feature extraction stages, each comprising four \nsk blocks. The first three blocks maintain constant output feature dimensions of 96, 192, 384, and 768 for the four stages, respectively. The final block in each stage increases the output feature dimensions to 128, 256, 512, and 1024, respectively.
Each block consists of two key components: a patch embedding module and a convolution module.
Within the patch embedding module, when it is located at the beginning of the last three stages, an average pooling layer with a kernel size and stride of 2 is first applied to downsample the feature map. Furthermore, for the modules in the first and last blocks of each stage, a linear projection is used to adjust the feature dimensions, followed by batch normalization to stabilize the feature distribution. For all other cases, the module performs an identity mapping.
The convolution module applies grouped $3 \times 3$ convolutions, where the number of groups is set to the output feature dimensions divided by 16. It is followed by batch normalization, Mish activation~\cite{misra2019mish}, and a $1\times 1$ convolution for feature fusion. The output is added back to the patch-embedded input via a residual connection.
After the four-stage feature extraction, we then apply GeM pooling~\cite{berman2019multigrain} to aggregate the representation along the temporal dimension, which is finally passed through a linear layer to produce the class probabilities.

\vpara{\tcn.}
\tcn\cite{donghao2024moderntcn} is a temporal convolutional architecture designed for time series data. The model is organized into three hierarchical stages, each containing an embedding layer and three \tcn blocks. Across these stages, the feature dimension progressively increases from 32 to 64 and then to 128. The embedding layer in the first stage differs from those in the subsequent stages. Specifically, it performs patch embedding independently for each channel using a one-dimensional convolutional layer with an output feature dimension of 32, a kernel size of 50, and a stride of 50. In contrast, the embedding layers at the beginning of the second and third stages apply linear projection operations to increase the feature dimension for each channel. Subsequently, the channel and feature dimensions are concatenated and fed into a \tcn block. Each \tcn block comprises three main components: a convolutional module, a feature-wise feed-forward network (FFN), and a channel-wise FFN. The convolution module employs large-kernel depthwise convolutions with kernel sizes of 21, 17, and 13 in the first, second, and third stages, respectively, complemented by auxiliary small-kernel branches with a kernel size of 5 to enhance local feature modeling. The feature-wise FFN groups convolutions by the number of channels $C$, while the channel-wise FFN groups them by the number of features. Both networks expand the hidden dimension four times via a linear projection, GELU activation~\cite{hendrycks2016gaussian}, dropout, and residual connections, before projecting it back to the original size. Finally, the extracted features are averaged along the temporal dimension, and the channel and feature dimensions are concatenated and fed into a linear classification head for discrete category prediction. 

\vpara{\medformer.}
\medformer~\cite{wang2024medformer} is a multi-granularity patching Transformer architecture specifically developed for medical time-series classification. The model initiates processing with a multi-scale patch embedding module that segments the input signals into non-overlapping temporal patches of lengths 5, 10, 20, and 50, thereby capturing both fine- and coarse-grained temporal dynamics. Each patch is projected into a 384-dimensional embedding space using a set of cross-channel token embedding layers, to which learnable contextual tokens and sinusoidal positional encodings are added to preserve the channel and temporal order.
The embeddings derived from the various patch scales are then independently processed by six \medformer encoder blocks, each comprising intra-scale and inter-scale self-attention mechanisms as well as a FFN. The intra-scale self-attention is applied separately to the patch sequences within each temporal scale, enabling the model to capture scale-specific temporal dependencies. Conversely, the inter-scale self-attention operates on a set of router tokens, which are obtained by extracting the last token from each intra-scale output. These router tokens interact through self-attention to facilitate the exchange of information across different temporal scales. The updated router representations then replace the original tokens in each scale sequence. The FFN expands the hidden dimension four times via a linear projection, ReLU activation, dropout, and layer normalization, before projecting it back to the original size. Residual skip connections are implemented around both attention and FFN to stabilize training. Finally, the router tokens from all scales are concatenated and passed through a linear projection layer to produce the class probabilities. 

\vpara{\gru.}
\gru~\cite{zinxira2024tlvmc} is a hierarchical recurrent neural network designed to capture temporal dependencies in time series data. The architecture begins with a linear embedding layer that projects the input into a 512-dimensional latent space, followed by layer normalization and a ReLU activation function. Subsequently, the model incorporates a stack of three residual bidirectional GRU blocks, each comprising a bidirectional GRU layer and a two-layer FFN. Within each block, the FFN initially expands the hidden representation to 1536 via a linear projection, followed by a ReLU activation, dropout, and layer normalization, and then projects the features back to the original hidden size using the same structure. Finally, the features are averaged along the temporal dimension and passed to a linear classification head for discrete category prediction.

\subsection*{Beam search}
\label{beam-search}
Beam search was used to decode the predicted probability distributions of the model on the initials and finals into candidate syllable sequences. Specifically, inference was first performed on the test sentences using the trained brain decoder to obtain the probability distributions of the initial and final components for each character. These probabilities were then organized according to the initial–final structure to construct a probability matrix that represents the likelihoods of different initial–final combinations for each character. In addition, a separator symbol (``+") with a fixed probability of 1 was inserted between consecutive initial–final pairs, ensuring that the beam search treated each complete syllable as a distinct decoding unit.

A lexicon-constrained beam search was subsequently applied to generate valid syllable sequences from the probability matrix. The lexicon was constructed by segmenting the decoding corpora into words and converting each word into its corresponding syllable sequence, which was organized as a prefix tree for efficient lookup. During decoding, the beam width was set to 100, a top‑k filter of 50 was applied, and the 20 most probable candidate sequences were retained. At each step of the beam search, up to 100 hypotheses were maintained. These hypotheses were expanded by querying the prefix tree with the top-50 most probable syllables from the model’s output distribution, ensuring that only syllable sequences corresponding to valid prefixes of lexicon entries were considered. Finally, the 20 complete hypotheses with the highest cumulative log‑probabilities were selected, yielding the most likely decoding candidates.

\subsection*{Syllable-to-sentence decoding framework}
\label{syllable-to-sentence-decoding-framework}
After obtaining the beam search candidates, we employed an LLM to generate the correct sentences. We observed that directly using the LLM for inference yielded suboptimal results, especially with small- or medium-scale LLMs, which sometimes produced outputs completely unrelated to the ground truth. Although very large commercial LLMs can generate reasonable results, their inference demands significant computational resources.
Moreover, due to the extreme sensitivity of clinical data, hospital data storage environments are typically isolated from external networks. As a result, decoding models must be deployed locally in practical decoding scenarios, a requirement that commercial LLMs often struggle to meet.
Therefore, our goal was to enable small-scale LLMs to effectively handle this task.

Despite selecting the top-20 candidates based on beam search scores, the quality of these candidates was highly variable in practice, often including many samples with extremely high error rates that could mislead the LLM. 
To address this issue, we decomposed the task into two simpler subtasks. First, the LLM selects the three best candidates from all beam search outputs. Then, it infers the correct sentence based on these three selected candidates.
There are two reasons for choosing three candidates: first, the best candidate is often not unique; second, the candidates may complement each other. For example, a certain syllable may be incorrect in the first candidate but correct in the second. Allowing the LLM to consider multiple high-quality candidates enables it to integrate complementary information and produce a more accurate result.
In our implementation, we utilized a publicly available Chinese-oriented LLM, \llmbase\cite{qwen2025qwen25technicalreport}, as the base model. 

\vpara{Vocabulary expansion.}
There are a total of 416 toneless syllables in Hanyu Pinyin~\cite{PinyinInfoHanyuRomanization}, of which the vocabulary of \llmbase covers 202 (Supplementary 5). Therefore, we extended the vocabulary to include all syllables. 

\vpara{Corpus for post-training.}
Because our recorded training set contains only individual-character data, post-training the LLM requires constructing synthetic data from publicly available datasets.
We built our corpus from two publicly available datasets: NLPCC18~\cite{nlpcc2018} and SIGHAN15~\cite{sighan2015}, which consist of relatively simple Chinese sentences. We filtered these datasets by retaining only sentences composed exclusively of Chinese characters. The filtered sentences and their corresponding toneless syllable sequences formed the post-training corpus. 

\vpara{Post-training task 1: translation.}
The translation task involves translating toneless syllable sequences into corresponding Chinese sentences, which serves as a preliminary task for the post-training process. We designed this task to allow the LLM to build a connection between unfamiliar knowledge (syllable sequences) and familiar knowledge (the Chinese language). We denote the model fine-tuned with this task as \llmtranslate.

\vpara{Post-training task 2: listwise ranking.}
The listwise ranking task requires the LLM to select the three candidates closest to the correct syllable sequence from a set of 20. 
Because the post-training corpus only contained the correct syllable sequences of the sentences, we needed to construct candidate syllable sequences from the original sequences to simulate those obtained from beam search.

We describe the steps that construct one candidate from a correct syllable sequence with $n$ syllables.
First, we decided on the error rate $r$ of the candidate. There were three error rate ranges: (0, 0.3), (0.3, 0.6), and (0.6, 0.9), from which one range was randomly selected according to probabilities drawn from a Dirichlet distribution with weights of 2, 2, and 1. The error rate $r$ was uniformly sampled within the range. In addition, if the selected range is (0, 0.3), the error rate $r$ was set to 0 with a probability of 10\%.
Second, we performed $\lfloor n\times r\rfloor$ random replacements. In each replacement, we randomly selected a syllable. If the selected syllable corresponded to a single Chinese character in the original sentence, it was replaced with another randomly chosen syllable. If the syllable corresponded to a character that formed a word together with other characters, then the entire word was replaced, ensuring that the syllable sequence of the substitute word differed from that of the original word by exactly one syllable. This replacement strategy enables the distribution of the generated data to more closely match the distribution of actual beam search outputs. Since different replacement operations might affect the same syllable, we recalculated the error rate of the candidate relative to the original syllable sequence after all replacement steps were completed to make corrections.

By independently repeating the above steps 20 times, we obtained a candidate set for each sample.
The fine-tuning of listwise ranking was based on \llmtranslate. When training the model with the constructed data, we randomly permuted the input candidates and asked the model to find out three candidates with the lowest error rates. After training, we obtained \llmrank.

\vpara{Post-training task 3: correction.}
In the correction task, the model was trained to predict the correct Chinese sentence based on the selected three candidate syllable sequences. We employed \llmrank to infer the data from the listwise ranking task, deriving the training data for correction. Using the previously trained model for inference, rather than directly selecting the three candidates with the lowest error rates, was motivated by the fact that the model-inferred candidates may not always be optimal. Incorporating relatively suboptimal candidates helps enhance the robustness of the model. The correction task was fine-tuned based on \llmrank, and we denote our correction model as \llmcorrect.

\vpara{Two-stage inference.}
We employed \llmcorrect as our inference model. Firstly, the top-20 candidates output from the beam search process were fed into the model to select the top three. Then, the selected candidates were input into the model again to generate the sentence.

\vpara{Ablation study.}
In the ablation study, we constructed a series of ablated models to evaluate the technical contributions of our syllable-to-sentence decoding framework. The IME-style decoder, implemented using the Python Pinyin2Hanzi package, relies on lexicon- and language-model–driven search for syllable-to-sentence conversion. For the direct correction approach, both the top-1 and top-3 candidate settings utilized inference via \llmcorrect to control variables consistently. Since \llmcorrect does not support direct correction of 20 candidates simultaneously, we trained a separate post-trained model based on \llmtranslate for the top-20 candidate setting. Ablations of the vocabulary expansion and translation post-training components were performed by removing these modules from the original framework.

\subsection*{Training and evaluation details}
In this section, we present the details for all the training and evaluation in our decoding framework.

\vpara{Data augmentation.}
In the classification of syllable components, we employed a comprehensive data augmentation pipeline comprising five techniques to enrich data diversity and enhance model robustness. 
First, a random shift was applied with a probability of 0.5, which shifted the input sequence randomly within ±10\% of its length, where positive and negative values indicate forward and backward shifts, respectively.
Second, additive noise was applied with a probability of 0.1, where zero-mean Gaussian noise was adaptively scaled according to the input’s standard deviation to yield a signal-to-noise ratio that was randomly sampled between 15 and 30 dB.
Third, channel masking was applied with a probability of 0.5, where each channel had an independent masking chance of 0.2. 
Fourth, time masking was applied with a probability of 0.5, in which four consecutive temporal segments, each covering 5\% of the total sequence length, were masked along the time axis. 
Finally, mixup~\cite{zhang2017mixup} was used with a probability of 0.5, where two samples and their corresponding labels were linearly combined using a mixing coefficient $\lambda$ drawn from a beta distribution $\alpha = 0.4$, thus smoothing the decision boundaries and improving generalization.

\vpara{Brain decoder training.}
All models were initialized with a random seed of 42 to ensure reproducibility. We performed training with a batch size of 64 and evaluation with a batch size of 128, using binary cross-entropy (BCE) loss. Optimization followed the Muon algorithm~\cite{liu2025muon}, with a weight decay coefficient of 0.05 serving as regularization. During the warm-up stage, which covered 10\% of the total training steps, the learning rate was linearly increased to $3 \times 10^{-4}$. After that, a cosine decay schedule was applied, maintaining the minimum learning rate at zero to promote smoother convergence and more stable optimization dynamics. 

For channel contribution analysis, we used only each participant's training data, which were further split into training and validation sets at a 4:1 ratio. We trained the model for 100 epochs and retained the checkpoint corresponding to the best validation performance. 
For syllable component decoding of four brain decoders, models were trained on the full training set for 500 epochs. 
To enhance model generalization, we applied Stochastic Weight Averaging (SWA)~\cite{izmailov2018averaging} during training, from the 250th to the 500th epoch. 
During evaluation, Test-Time Augmentation (TTA)~\cite{shanmugam2021better} was employed to improve prediction stability. For each test sample, the original input was retained along with two additional versions generated via random shifts. The model performed inference on all three variants, and their output logits were averaged to produce the final prediction. Moreover, to ensure evaluation reliability and reduce randomness, we performed inference under ten different random seeds, and the averaged results were reported.

\vpara{LLM post-training.}
For all post-training tasks of the LLM, we performed supervised fine-tuning (SFT) using LoRA~\cite{hu2022lora} with a rank of $r=16$ and a scaling factor of $\alpha=32$. Each task was trained for one epoch and optimized with the AdamW optimizer. We adopted a linear warm-up over 5\% of the training steps to reach a peak learning rate of $5 \times 10^{-5}$, followed by a cosine scheduler that decayed the learning rate to zero. To further improve computational efficiency, BFloat16 precision~\cite{kalamkar2019study} and FlashAttention‑2~\cite{dao2023flashattention} were employed throughout the fine‑tuning process. The detailed tuning instructions are summarized in Supplementary 6.

\subsection*{Statistical analysis}
\label{statistical-analysis}
In this section, we describe the statistical analyses used for all experimental results reported in this study. Different statistical tests were applied depending on the comparison setting, as detailed below.

For comparisons between speech production and speech perception within the same set of highly responsive channels (Fig.~\ref{fig:Fig2}c), performance differences between speaking and listening tasks were assessed using two-sided Wilcoxon signed-rank tests.
For comparisons of decoding performance between the left and right hemispheres (Fig.~\ref{fig:Fig2}f), effect sizes were quantified using Cliff’s delta, which is appropriate for non-parametric comparisons of independent samples.
For initial and final classification performance, one-sided Wilcoxon signed-rank tests were conducted to assess whether each participant’s decoding accuracy significantly exceeded chance level. Comparisons between different brain decoders (Fig.~\ref{fig:Fig3}a) were conducted across all participants using two-sided Wilcoxon signed-rank tests. 
To assess the relationship between the improvement ratio in the EMP and that in the initial–final classification performance (Fig.~\ref{fig:Fig4}b), we applied a one-sided Wilcoxon signed-rank test to examine whether the distribution of their paired differences was significantly greater than zero.
For sentence-level decoding performance (Fig.~\ref{fig:Fig4}c,d), comparisons among different LLMs, as well as between our proposed LLM and its ablated variants, were performed across all participants using two-sided Wilcoxon signed-rank tests.

\section*{Data availability}
Because of ethical restrictions, the dataset cannot be publicly archived. However, it is available on request from the senior corresponding author.
The source data accompanying this paper include all quantitative results reported in the manuscript, compiled into \texttt{Supplementary Tables.xlsx}, along with a \texttt{README.md} file that describes the contents of each sheet.

\section*{Code availability}
The source code supporting this study is made available with the paper.
Detailed implementation and usage information can be found in the accompanying \texttt{README.md} file.

\section*{Use of LLMs}
\label{Use of Large Language Models (LLMs)}
In addition to the LLMs that we post-trained and evaluated within our own experiments, we made limited use of publicly available LLMs during the preparation of this manuscript. Specifically, an LLM was used to assist with minor writing refinements, such as improving grammatical accuracy and textual clarity. In some cases, the LLM was also employed for small-scale code completion tasks, including generating boilerplate functions or suggesting minor syntax corrections. All core ideas, experimental designs, analyses, and conclusions were completely conceived, implemented, and validated by the authors.

\section*{Acknowledgments}
\label{acknowledgements}
This work was supported by the National Science and Technology Major Project (2025ZD0215100) and the National Natural Science Foundation of China (Nos. 32595492 and 82272116).
We also gratefully acknowledge the support of the iBRAIN (Intracranial Brain Recording/Activation/Inhibition Network) Data Alliance. The iBRAIN Alliance is a multi-center collaborative initiative dedicated to establishing a large-scale, standardized intracranial EEG database to advance research in brain-computer interfaces and clinical neuroscience.

\section*{Author Contributions}
\label{author-contributions}
Z.Y., Y.Y., and M.L. conceived the study. Z.Y., G.Z., and B.C. designed the data acquisition paradigms. L.C. and X.L. were responsible for clinical electrode implantation. Z.W. managed the clinical trials. B.C. conducted data acquisition and preprocessing. Y.X. assisted with clinical data acquisition. Y.M. supervised the overall data acquisition. Z.Y. and G.Z. designed the model framework. G.Z. implemented the model and performed validation experiments. Z.Y., B.C., and M.L. designed the analysis experiments. Z.Y. conducted the analysis experiments. Z.Y. and G.Z. wrote the manuscript. Y.Y. revised the manuscript. Y.Y. and M.L. supervised the project.

\section*{Competing Interests}
\label{competing-interests}
The authors declare no competing interests.



\begin{appendices}
\clearpage
\section{Extended Data}

\subsection{Overall spatial distribution of channel contributions}
\label{overall-spatial-distribution-of-channel-contributions}

\begin{figure}[htbp]
   \centering
   \includegraphics[width=0.7\textwidth]{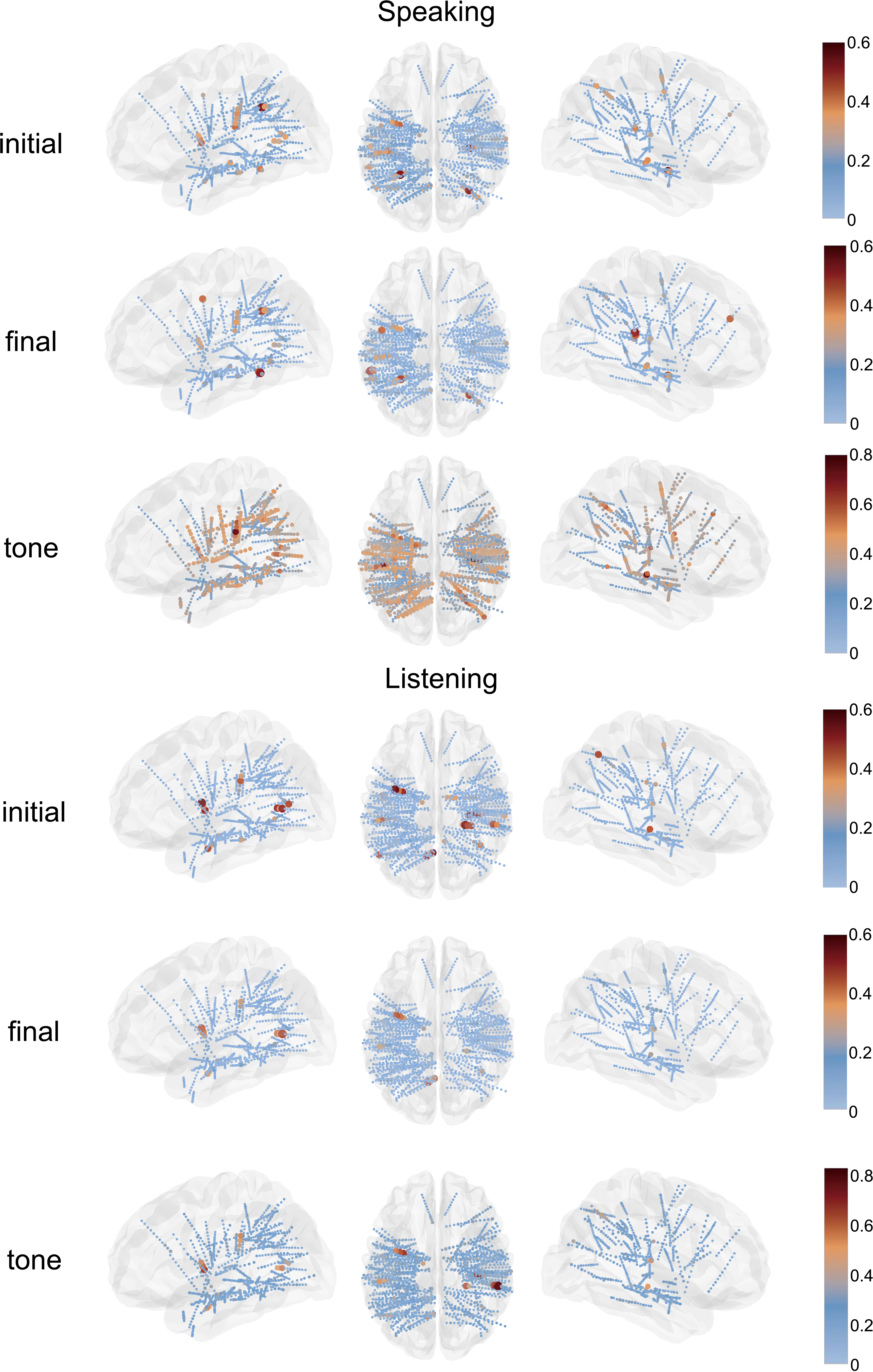}
   \caption{
   \textbf{Spatial distribution of sEEG electrodes and their contributions across the brain.}
   Electrodes from 12 subjects are visualized from three brain perspectives. Heatmaps indicate each channel’s F1 score in the speaking and listening tasks during initial, final, and tone decoding.
   } 
   \label{fig:ExtendA1}
\end{figure}

\clearpage

\subsection{Architectures of brain decoders}
\label{architectures of brain decoders}

\begin{figure}[htbp]
   \centering
   \includegraphics[width=0.9\textwidth]{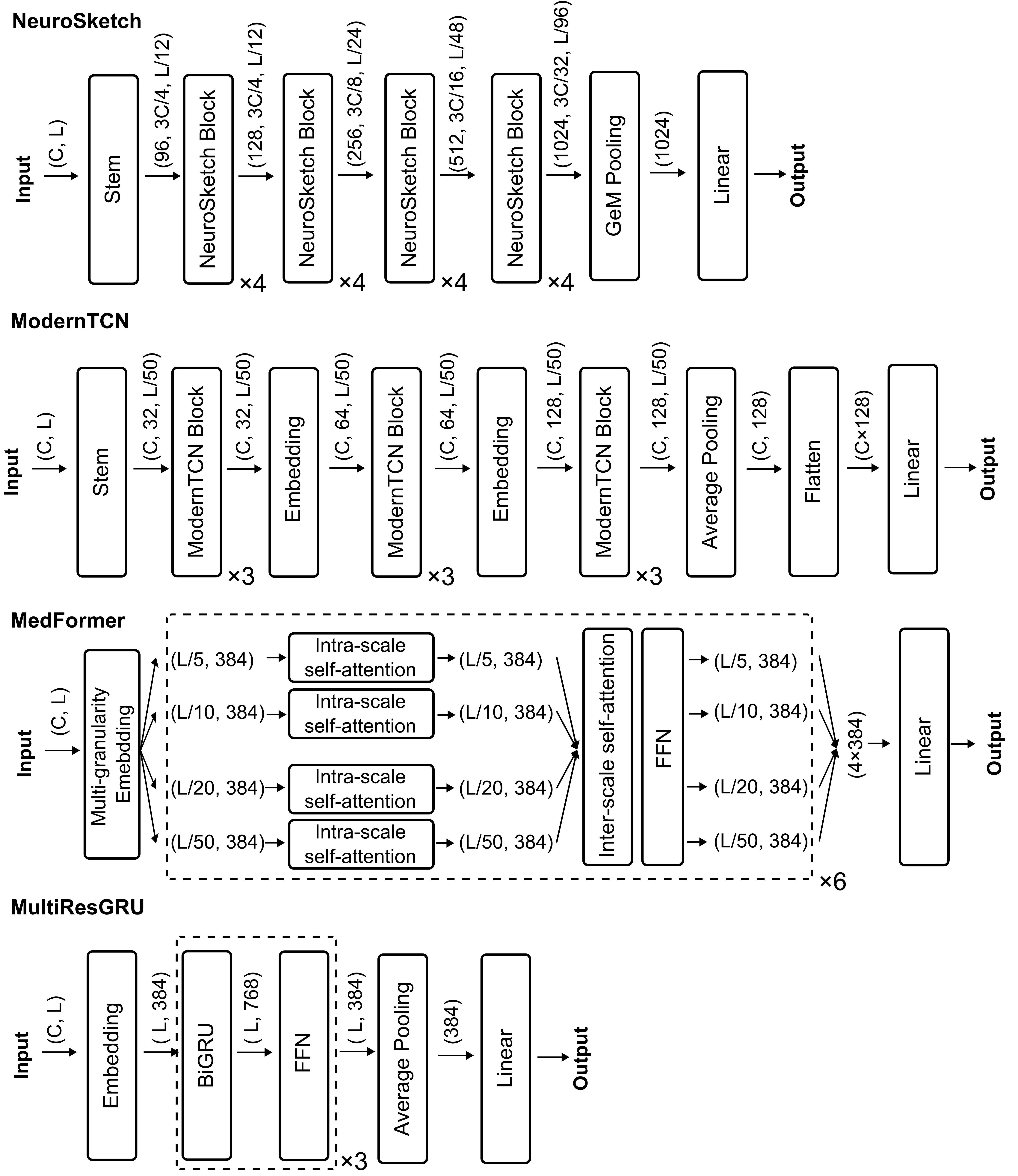}
   \caption{
   \textbf{Architectures of the four brain decoders.}
   The input to each brain decoder is an sEEG signal with $C$ channels and $L$ time steps. For each stage of the decoder, we indicate the number of output channels, the number of time steps, and the corresponding number of blocks.
   }
   \label{fig:ExtendA2}
\end{figure}

\clearpage




\subsection{F1 scores of initial-final classification}
\label{f1-scores-of-initial-final-classification}

\begin{figure}[htbp]
   \centering
   \includegraphics[width=1.0\textwidth]{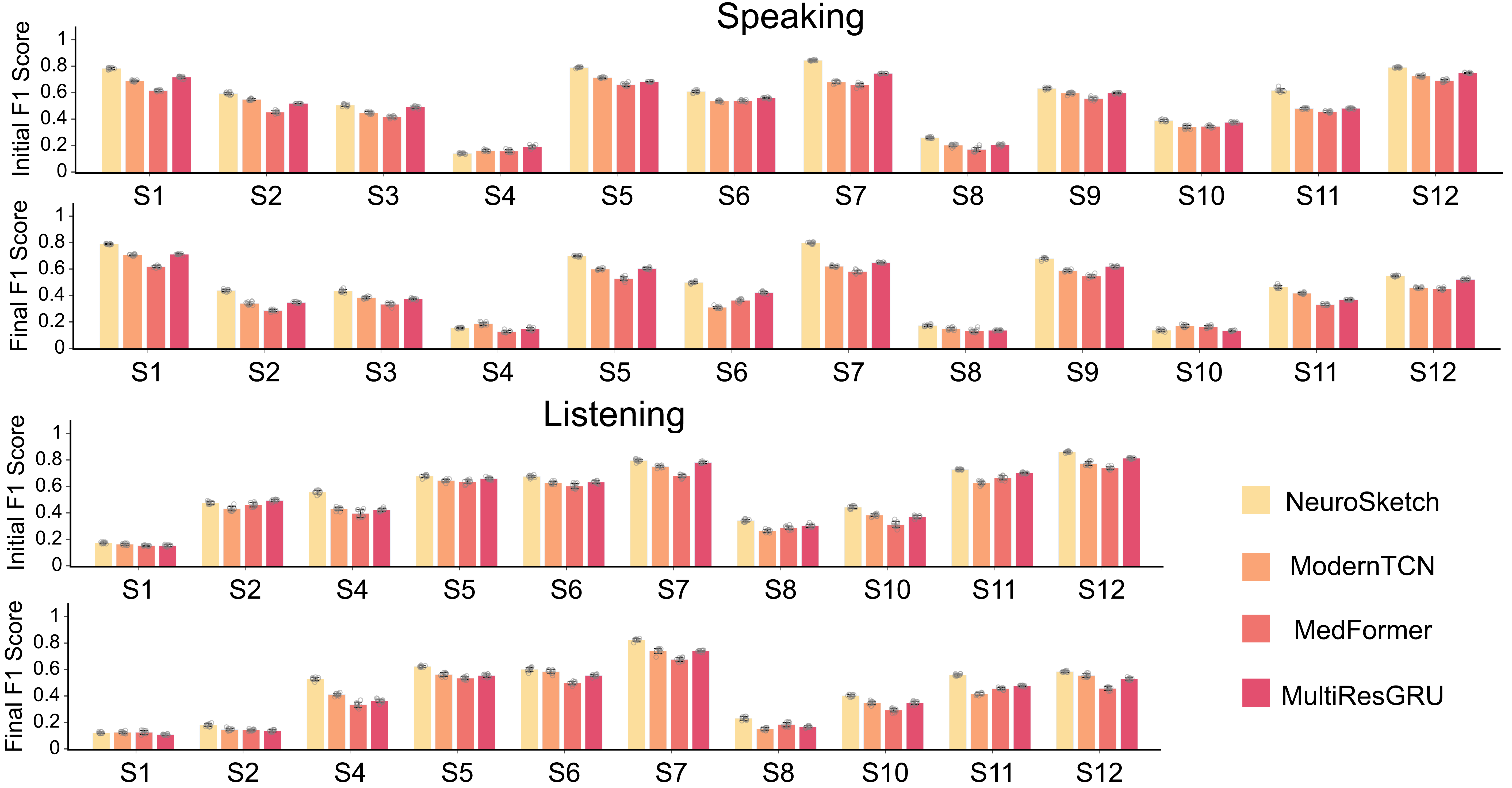}
   \caption{
    \textbf{Initial-final classification results}. 
   Bar plots illustrate the initial and final classification F1 scores achieved by four different brain decoders for speaking and listening tasks.
   } 
   \label{fig:ExtendA4}
\end{figure}

\clearpage

\subsection{F1 scores of tone classification}
\label{f1-scores-of-tone-classification}

\begin{figure}[htbp]
   \centering
   \includegraphics[width=1.0\textwidth]{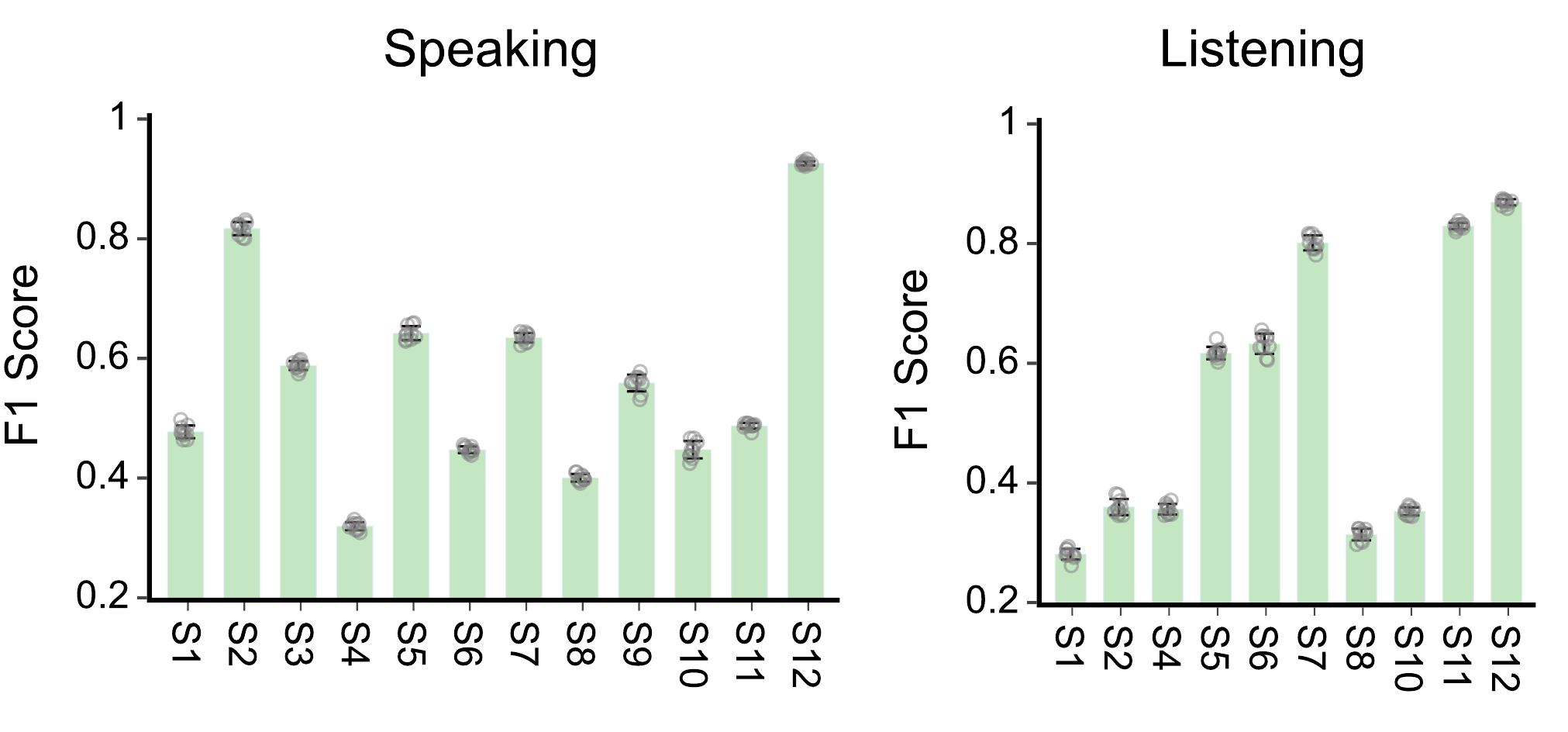}
   \caption{
    \textbf{Tone classification results}. 
   Bar plots illustrate the tone classification F1 scores achieved by four different brain decoders for speaking and listening tasks.
   } 
   \label{fig:ExtendA5}
\end{figure}

\clearpage

\end{appendices}

\end{document}